\documentclass[a4,14pt]{article}
\usepackage[margin=1in]{geometry}
\usepackage{subcaption}
\usepackage{amsmath}
\usepackage{hyperref}
\usepackage{amssymb}
\usepackage{braket}
\usepackage{upgreek}
\usepackage{graphicx}
\usepackage{float}
\begin{document}
\pagenumbering{arabic}
\title{Damped Neutrino Oscillations in a Conformal Coupling Model}
\author{H. Mohseni Sadjadi\footnote{mohsenisad@ut.ac.ir} and H. Yazdani Ahmadabadi\footnote{hossein.yazdani@ut.ac.ir}
\\ {\small Department of Physics, University of Tehran,}
\\ {\small P. O. B. 14395-547, Tehran 14399-55961, Iran}}
\maketitle
\begin{abstract}
Flavor transitions of Neutrinos with a nonstandard interaction are studied. A scalar field is conformally coupled to matter and neutrinos. This interaction alters the neutrino effective mass and its wavefunction leading to a damping factor, causing deficits in the probability densities and affecting the oscillation phase. As the matter density determines the scalar field's behavior, we also have an indirect matter density effect on the flavor conversion. We explain our results in the context of screening models and study the deficit in the total flux of electron-neutrinos produced in the Sun through the decay process and confront our results with observational data.
\end{abstract}

\section{Introduction}\label{sec1}
One of the interesting subjects in particle physics and cosmology is the physics of neutrinos, which has received a lot of attention, especially in the beyond standard model theories. In the standard model of particles, neutrinos were assumed to be massless. Massive neutrinos were first proposed theoretically in \cite{BP1,BP2} and after some decades finds severe attentions by detection of deficits in the number of electron neutrinos received from the Sun \cite{Sun1,Sun2}, and in the number of atmospheric muon neutrinos \cite{mu1,mu2}. We may explain the discrepancy between the expected number of neutrinos and the observation by considering the flavor state as mixed mass states. This gives rise to neutrinos flavors change via oscillation during their travels (in the atmosphere). Also, the influence of local electron density on electron-neutrino density gives rise to adiabatic flavor conversion (as in the Sun) through the MSW (Mikheyev-Smirnov-Wolfenstein) effect \cite{MSWWolf,MSWMS}. Apart from the standard weak interaction, one may consider nonstandard neutrino interactions (not included in the standard model) \cite{MSWWolf,NSI,NSI1} and investigate their influences on neutrino oscillations \cite{smir}. Nonstandard interaction of neutrinos and exotic fields has also attracted much attentions in the cosmology\cite{NSIC1,NSIC2,NSIC3,NSIC4,NSIC5,NSIC6}. These additional fields may be the dark energy describing the present cosmic acceleration of the Universe. Inspired by the similarity of the neutrino mass squared difference the scale of dark energy, mass varying neutrinos (whose mass depends on a light scalar field) were introduced in \cite{MV11,MV12,MV13} and used in \cite{MV1,MV2} to discuss the solar neutrino flavor survival probability. The dependence of the neutrino mass on the environment due to the interactions with dark sectors was also studied in \cite{Nel}, showing that the interaction alters the oscillation drastically, by affecting the neutrino mass.

In \cite{sad1,sad2,sam}, a model for the Universe cosmic acceleration \cite{amen} based on quintessence-neutrino interaction through a conformal coupling was proposed such that when the massive neutrinos became non-relativistic, by activating the quintessence, ignited the Universe acceleration. In this class of scalar-tensor dark energy models, the quintessence interacts with matter (including neutrinos)through a conformal coupling \cite{mat1,mat2,mat3}. This coupling may give rise to the screening effect as was studied in the chameleon \cite{Khoury:2003rn,Waterhouse:2006wv,Burrage:2017qrf,Tsujikawa:2009yf}and the symmetron\cite{Hinterbichler:2011ca, sad3} models. In these screening models, the scalar field's behavior is specified by the matter density such that in a dense region, they are screened. In this paper, we aim to study the effect of such a conformal coupling on the neutrinos' densities and oscillations. As the scalar field's behavior is specified by the matter density, we expect to encounter an MSW-like effect but caused by a nonstandard interaction, i.e., the conformal coupling. The interaction with the scalar field by modifying the neutrino effective mass and its density provides a mixed situation \cite{Mix} in which the neutrino deficit may be related to both the neutrino oscillation and neutrino decay. Such neutrino decay models as a second-order effect behind the solar neutrino problem can better fit solar neutrino data \cite{Aharmim:2018fme}, which can constrain the lifetime ($\tau_2$) of neutrino mass eigenstate $\nu_2$. In the decay formalism presented in \cite{Aharmim:2018fme,NeutLifTime}, since the actual neutrino masses ($m_i$) are unknown, the decay of mass eigenstate $\nu_2$ is completely explained by the ratio $\tau_i/m_i$. Fitting to all phases of $^8$B solar neutrino data by SNO (Sudbury Neutrino Observatory) \cite{Aharmim:2018fme} constrains ratio $\tau_2/m_2$ to be $>8.08 \times 10^{-5}$sec$/$eV at $90\%$ confidence level.

This paper is organized as follows: In sec.\ref{sec2} we discuss the ingredients of our model starting with a conformal transformation applied on the metric, which rescales relevant Dirac action governing the motion of neutrinos. Corrections to the neutrino mass and wavefunction related to the curved spacetime in the presence of a scalar field are explicitly calculated in this section. We then give an analytical solution to the mass-varying Dirac equation in subsec.\ref{sub21} and show how the scalar-matter coupling function will change the neutrino quantum mechanical phase. The damped two-flavor transition is the subject of the subsection\ref{sub22} giving the probabilities for both chameleon and symmetron cases. After that, we will extend damped neutrino transitions to three flavors and discuss the possibility of neutrino decay and deficit in the total probability in \ref{sub23}. Interactions with matter such as the Sun, for instance, have a dramatic effect on the flavor conversion, MSW (Mikheyev-Smirnov-Wolfenstein) effect, presented in the subsection\ref{sub24}. We will see that the LMA (large mixing angle) solutions have the most apparent effect on neutrino decay. Finally, we summarize and present our results in sec.\ref{sec3} through some numerical examples. A comparison of our results to the SNO \cite{Sun2}, SNO + SK (Super-Kamiokande) \cite{Zyla:2020zbs} and also to the Borexino \cite{Agostini:2018uly} data for the MSW-LMA survival probability will be done.

Throughout this paper we use units $\hbar=c=1$ and metric signature $(-,+,+,+)$.
\section{Conformal coupling and neutrino oscillation}\label{sec2}
Our model is specified by the following action including a scalar field conformally coupled to matter
\begin{eqnarray}\label{eqn1}
S= \int d^4x \sqrt{-g}\bigg[\frac{M_{p}^2}{2}R - \frac{1}{2} g^{\mu\nu} \partial_\mu \phi \partial_\nu \phi - V(\phi) \bigg] + \int d^4x  \mathcal{L}_m \big(\Psi_i , \tilde{g}_{\mu\nu}\big),
\end{eqnarray}
where $g$ is the determinant of the metric, $g_{\mu\nu}$, $M_{p}$ is the reduced Planck mass, $R$ is the Ricci scalar and $\mathcal{L}_m$ is the Lagrangian density of other components generally denoted by $\Psi_i$. $\tilde{g}_{\mu\nu}$ is related to the metric $g_{\mu\nu}$, by\cite{Faraoni:1998qx,Carneiro:2004rt,Bean}
\begin{eqnarray}\label{eqn2}
\tilde{g}_{\mu\nu} = A^2(\phi) g_{\mu\nu}.
\end{eqnarray}
The conformal factor $A(\phi)$ is a function of $\phi$. Various screening models are resulted by different choices of the coupling and potential functions, e.g. the chameleon model corresponds generally to the choice $A(\phi) \equiv \exp[{\frac{1}{M_p} \int \beta(\phi) d\phi}]$ where $\beta(\phi)$ is a field-dependent coupling parameter. A simple case corresponds to a constant value of $\beta \sim 1$. In the symmetron model a quadratic coupling function such as $A(\phi) \equiv 1 + \frac{\phi^2(r)}{2M^2}$  is chosen to respect $\mathbb{Z}_2$ symmetry. In the Appendix \ref{app1}, we will review these two screening mechanisms and point out the required relations for our discussion.
If we want to respect the weak equivalence principle (WEP), we must take a universal $A(\phi)$  \cite{Khoury:2003rn}; otherwise, we may have different $A_i(\phi)$ corresponding to different $\Psi_i$.

To compute the probability of neutrino flavor transition, we study the neutrino equation of motion. The neutrino action is
\begin{eqnarray}\label{eqn15}
S_n=\int d^4x \sqrt{-\tilde{g}}~\bar{\psi}(x)\left(i\tilde{\gamma}^\mu \tilde{D}_\mu-m\right)\psi(x).
\end{eqnarray}
To obtain the covariant derivative we proceed as follows:
The \textit{vierbein} or \textit{tetrad}, $\epsilon_a^\mu$, are
\begin{eqnarray}\label{eqn5}
\tilde{\epsilon}_{\mu}^a = A(\phi)\epsilon_{\mu}^a,~~~~~~\tilde{\epsilon}_a^{\mu} = A^{-1}(\phi) \epsilon_a^{\mu}.
\end{eqnarray}
The Dirac $\gamma$-matrices and the Christoffel connections $\Gamma_{\mu\nu}^{\lambda}$ are given by
\begin{eqnarray}\label{eqn6}
\tilde{\gamma}^\mu = A^{-1}(\phi) \gamma^\mu,
\end{eqnarray}
and
\begin{eqnarray}\label{eqn7}
\tilde{\Gamma}_{\mu\nu}^{\lambda} = \Gamma_{\mu\nu}^{\lambda} - A^{-1}(\phi) A^{,\lambda}(\phi)g_{\mu\nu} + A^{-1}(\phi) A_{,\mu} (\phi) \delta_{\nu}^{\lambda} + A^{-1}(\phi) A_{,\nu}(\phi) \delta_{\mu}^{\lambda}.
\end{eqnarray}
respectively. The spin connection is
\begin{eqnarray}\label{eqn8}
\tilde{\omega}_{\mu}^{ab}=\omega_{\mu}^{ab} - A^{-1}(\phi) A_{,\nu}(\phi) \epsilon^{a\nu}\epsilon_\mu^b + A^{-1}(\phi) A_{,\nu}(\phi) \epsilon^{b\nu}\epsilon_\mu^a,
\end{eqnarray}
in which $\omega_{\mu}^{ab}=-\epsilon^{b\nu}({\partial_\mu}\epsilon_{\nu}^a-\Gamma_{\mu\nu}^{\lambda}\epsilon_{\lambda}^a)$.
So the covariant derivative is
\begin{eqnarray}\label{eqn9}
\begin{split}
&\tilde{D}_\mu=\partial_{\mu}+i \tilde{\omega}_\mu^{ab} \Sigma_{ab}\\&~~~~=D_\mu + iA^{-1}(\phi) A_{,\nu}(\phi) \big(\epsilon^{b\nu}\epsilon_\mu^a - \epsilon^{a\nu}\epsilon_\mu^b\big)\Sigma_{ab}.
\end{split}
\end{eqnarray}
The commutation of  $\gamma$-matrices, is unchanged under the conformal transformation:
\begin{eqnarray}\label{eqn10}
\begin{split}
&\tilde{\Sigma}_{ab}=\dfrac{-i}{8}[\tilde{\gamma}_a,\tilde{\gamma}_b]=[\tilde{\epsilon}_{a\mu}\tilde{\gamma}^\mu,\tilde{\epsilon}_{b\nu}\tilde{\gamma}^\nu]\\&~~~~~=\dfrac{-i}{8}[A(\phi) A^{-1}(\phi) \epsilon_{a\mu}\gamma^\mu , A(\phi) A^{-1}(\phi) \epsilon_{b\nu}\gamma^\nu]\\&~~~~~=\dfrac{-i}{8}[\gamma_a,\gamma_b]=\Sigma_{ab}.
\end{split}
\end{eqnarray}
The second term in (\ref{eqn9}) can be written in a more familiar form
\begin{eqnarray}\label{eqn11}
\big(\epsilon^{b\nu} \epsilon_\mu^a-\epsilon^{av}\epsilon_\mu^b\big)\Sigma_{ab}=\dfrac{-i}{4}\big(\gamma_\mu \gamma^\nu-\gamma^\nu \gamma_\mu\big).
\end{eqnarray}
Therefore the covariant derivative $\tilde{D}_\mu$ can finally be written as
\begin{eqnarray}\label{eqn12}
\tilde{D}_\mu=D_\mu+\frac{1}{4} A^{-1}(\phi) A_{,\nu}(\phi) \left(\gamma_\mu \gamma^\nu-\gamma^\nu \gamma_\mu\right),
\end{eqnarray}
leading to
\begin{eqnarray}\label{eqn13}
\tilde{\gamma}^\mu \tilde{D}_\mu = A^{-1}(\phi) \gamma^\mu D_\mu + \dfrac{3}{2} A^{-2}(\phi) A_{,\mu}(\phi) \gamma^\mu.
\end{eqnarray}
One can use this relation to show that
\begin{equation}
\int d^4x \sqrt{-\tilde{g}}~\bar{\psi}(x)\left(i\tilde{\gamma}^\mu \tilde{D}_\mu-m\right)\psi(x)=\int d^4x \sqrt{g}\bar{\psi^\prime}(x)\left(i\gamma^\mu D_\mu-m^\prime\right)\psi^\prime(x),
\end{equation}
provided that
\begin{eqnarray}\label{eqn17}
m^\prime (x) = m A(\phi),
\end{eqnarray}
and
\begin{equation}\label{eqn18}
\psi^\prime(x) = \psi(x) A^{\frac{3}{2}}(\phi).
\end{equation}

We fix $g_{\mu\nu}$ as the Minkowski metric $\eta_{\mu\nu}$. So the covariant derivative $D_{\mu}$ simply becomes the partial derivative $\partial_{\mu}$.
\subsection{Neutrino oscillation}\label{sub21}
Neutrino oscillation or mixing can be explained in terms of the relationship between flavor eigenstates ($\nu_e,\nu_\mu,\nu_\tau$) and mass eigenstates ($\nu_1,\nu_2,\nu_3$). This implies that we can write each flavor eigenstate as a superposition of three mass eigenstates. Mixing is given by
\begin{eqnarray}\label{eqn19}
\ket{\upnu_i} = \sum_{\alpha} U_{\alpha i} \ket{\upnu_\alpha},
\end{eqnarray}
where $ \ket{\upnu_i}=
\begin{pmatrix}
\nu_1\\
\nu_2\\
\nu_3
\end{pmatrix}
$ and
$ \ket{\upnu_\alpha}=
\begin{pmatrix}
\nu_e\\
\nu_\mu\\
\nu_\tau
\end{pmatrix}
$
are mass and flavor eigenstates, respectively, and $U$, which is a unitary mixing matrix, called PMNS (Pontecorvo–Maki–Nakagawa–Sakata) matrix, can be written as
\begin{eqnarray}\label{eqn20}
U =
\begin{pmatrix}
U_{e 1} & U_{e 2} & U_{e 3} \\
U_{\mu 1} & U_{\mu 2} & U_{\mu 3} \\
U_{\tau 1} & U_{\tau 2} & U_{\tau 3} \\
\end{pmatrix}
.
\end{eqnarray}

Dirac Lagrangian density describing both right-handed and left-handed neutrinos including the mass and dynamical terms is given by \cite{Cottingham:2007zz}
\begin{eqnarray}\label{eqn21}
\begin{split}
&\mathcal{L^\prime} = \mathcal{L^\prime}_{mass} + \mathcal{L^\prime}_{dyn.} = - \sum_{\alpha , \beta} \upnu_{\alpha L}^{\prime \dagger} m^\prime_{\alpha \beta} \upnu^\prime_{\beta R} - \sum_{\alpha , \beta} \upnu_{\alpha R}^{\prime \dagger} m_{\beta \alpha}^{\prime *} \upnu^\prime_{\beta L} \\&~~~~~~~~~~~~~~~~~~~~~~~~~~~~~~+ \sum_{\alpha} i [\upnu_{\alpha L}^{\prime \dagger} \sigma_L^\mu \partial_\mu \upnu^\prime_{\alpha L} + \upnu_{\alpha R}^{\prime \dagger} \sigma_R^{\mu} \partial_\mu \upnu^\prime_{\alpha R}].
\end{split}
\end{eqnarray}
where $m^\prime_{\alpha \beta}$'s are components of a $3 \times 3$ mass matrix in the flavor basis, $\upnu^\prime_{\alpha (R,L)}$ are 2-spinors implying right-handed and left-handed neutrinos of flavor $\alpha$. In addition, $\sigma_R^{\mu} \equiv (\sigma^{0},\sigma^{1} , \sigma^{2} , \sigma^{3})$ and $\sigma_L^\mu \equiv (\sigma^{0},-\sigma^{1} , -\sigma^{2} , -\sigma^{3})$ are Pauli matrices for right-handed and left-handed spinors, respectively. We also note that all primes in the formulas refer to the rescaled relations.
\\Now, by varying this Lagrangian with respect to the spinors, equations of motion can be obtained. For the first one, we have
\begin{eqnarray}\label{eqn22}
i \sigma_L^{\mu} \partial_{\mu} \upnu^\prime_{\alpha L} - m^\prime_{\alpha \beta} \upnu^\prime_{\beta R} = 0,
\end{eqnarray}
and the second equation is given by
\begin{eqnarray}\label{eqn23}
i \sigma_R^{\mu} \partial_{\mu} \upnu^\prime_{\alpha R} - m_{\beta \alpha}^{\prime *} \upnu^\prime_{\beta L} = 0.
\end{eqnarray}
Generally, the left-handed and right-handed spinors can be defined by
\begin{eqnarray}\label{eqn24}
\upnu^\prime_{\alpha L}(r,t) = e^{-iE(t-t_0)}e^{+iE(r-r_0)} f_{\alpha}(r)
\begin{pmatrix}
0\\
1\\
\end{pmatrix}
,
\end{eqnarray}
\begin{eqnarray}\label{eqn25}
\upnu^\prime_{\alpha R}(r,t) = e^{-iE(t-t_0)}e^{+iE(r-r_0)} g_{\alpha}(r)
\begin{pmatrix}
0\\
1\\
\end{pmatrix}.
\end{eqnarray}
Substituting these relations into Eqs.(\ref{eqn22}) and (\ref{eqn23}), we obtain
\begin{eqnarray}\label{eqn26}
i \frac{d}{dr}f_\alpha (r) - m^\prime_{\alpha  \beta} g_{\beta} (r) = 0,
\end{eqnarray}
\begin{eqnarray}\label{eqn27}
\bigg[2E - i \frac{d}{dr}\bigg]g_\gamma (r) - m_{\alpha \gamma}^{\prime *} f_\alpha (r) = 0.
\end{eqnarray}
Notice that derivatives in relations above are all taken with respect to the radial coordinate $r$, because we have only considered the radial propagation in our model to solve Eqs.(\ref{eqn26}) and (\ref{eqn27}). For neutrinos masses much less than their energies, the second term in the square brackets in Eq.(\ref{eqn27})is negligible with respect to the first one. By applying this approximation, we obtain
\begin{eqnarray}\label{eqn28}
g_\gamma (r) = \frac{m_{\alpha \gamma}^{\prime *} f_\alpha (r)}{2E}
\end{eqnarray}
and by putting this into Eq.(\ref{eqn26}), we have
\begin{eqnarray}\label{eqn29}
i\frac{d}{dr} f_\beta (r) - \frac{m^\prime_{\beta \gamma} m_{\alpha \gamma}^{\prime *}}{2E} f_\alpha (r) = 0,
\end{eqnarray}
where $m^\prime_{\beta \gamma} = U^{L*}_{\beta i} U^R_{\gamma i} m^\prime_i$, $m_{\alpha \gamma}^{\prime *} = U^{L}_{\alpha j} U^{R*}_{\gamma j} m^\prime_j$ and we know $U^{L(R)} U^{L(R)\dagger} =1$. Dropping superscripts $L$, Eq.(\ref{eqn29}) leads us to the following equation
\begin{eqnarray}\label{eqn30}
i\frac{df_\beta (r)}{dr} = U_{\beta i}^* U_{\alpha i} \bigg(\frac{m_i^{\prime 2}}{2E}\bigg) f_\alpha (r).
\end{eqnarray}
By using $f_i(r) = U_{\alpha i} f_{\alpha} (r)$ and also Eq.(\ref{eqn30}) we obtain
\begin{eqnarray}\label{eqn31}
i\frac{df_i (r)}{dr} = \bigg(\frac{m_i^{\prime 2} (r)}{2E}\bigg) f_i (r).
\end{eqnarray}
This equation can be easily solved by
\begin{eqnarray}\label{eqn32}
f_i(r) = f_i(r_0) \exp\bigg[{-i\int_{r_0}^{r} \frac{m_i^{\prime 2}(r)}{2E} dr}\bigg].
\end{eqnarray}
Therefore, according to Eq.(\ref{eqn24}), the normalized spacetime part of the neutrino wavefuction denoted by $\Psi(r,t)$ satisfying the initial condition is given by (see \cite{MohseniSadjadi:2017jne,Cardall:1996cd,Visinelli:2015uva,Buoninfante:2019der,Chakraborty:2015vla})
\begin{eqnarray}\label{me}
\Psi_i(r,t)&=& e^{-iE[(t-t_0) - (r-r_0)]}e^{-i\varphi_{i}(r)} \mathcal{D}^{\frac{1}{2}}_{ii} (r) \Psi_i (r_0,t_0)\nonumber \\
&&\equiv \mathcal{F}_i(r,t)\Psi_i (r_0,t_0)
\end{eqnarray}
where
\begin{eqnarray}\label{eqn35}
\mathcal{D}_{ij}(r) \equiv [A_i(\phi) A_j(\phi) A_{0i}^{-1} (\phi) A_{0j}^{-1} (\phi)]^{-\frac{3}{2}}.
\end{eqnarray}
As we will see depending on the model, $\mathcal{D}$ may be a damping or an enhancing factor. We will note it by $\mathcal{D}$ factor.
\begin{eqnarray}\label{eqn36}
\varphi_{i}(r) \equiv \frac{m_i^2}{2E} \int_{r_0}^r A_i^2[\phi(r^\prime)] dr^\prime
\end{eqnarray}
is the phase of the oscillation. Note that in (\ref{eqn35}) and(\ref{eqn36}), different $A$ for different neutrinos are generally assumed. By $ A_{0i}$ we mean the value of $A_i$ at the initial point $(r_0,t_0)$. Notice that the initial value of the function $\mathcal{F}_i(r,t)$ is clearly equal to one, i.e. $\mathcal{F}_i(r_0,t_0) = 1$. In what follows, the above results will be used to calculate the oscillation probability.
\subsection{Two-flavor neutrino transitions}\label{sub22}
We first consider two neutrino flavors: $\ket{\nu_e}$ and $\ket{\nu_\mu}$. Using (\ref{me}), the general state of a propagating neutrino in the two-flavor basis is given by
\begin{eqnarray}\label{eqn37}
\ket{\upnu(r,t)} = \cos\theta~\Psi_1(r,t) \ket{\nu_1} + \sin\theta ~\Psi_2(r,t) \ket{\nu_2},
\end{eqnarray}
 We have
\begin{eqnarray}\label{eqn38}
\ket{\upnu(r_0,t_0)}= \cos\theta \ket{\nu_1}\Psi_1(r_0,t_0) + \sin\theta \ket{\nu_2}\Psi_2(r_0,t_0).
\end{eqnarray}
where we have used $\mathcal{F}_i(r_0,t_0) = 1$. Choosing the initial condition as $\Psi_1(r_0,t_0)=\Psi_2(r_0,t_0)=:\Psi_0$, we find
\begin{equation}
\ket{\upnu(r_0,t_0)}=\Psi_0 \ket{\upnu_e}
\end{equation}
where we have used $\ket{\upnu_e}=\cos\theta \ket{\nu_1} + \sin\theta \ket{\nu_2}$.

At the moment, we pay attention to the probability of neutrino oscillation for the case of two-flavor neutrinos, e.g. $\nu_e$ and $\nu_\mu$. As we know, the relation between mass and flavor eigenstates can be described by a $2 \times 2$ mixing matrix as follows
\begin{eqnarray}\label{eqn39}
\begin{pmatrix}
\nu_1\\
\nu_2\\
\end{pmatrix}=
\begin{pmatrix}
\cos\theta & -\sin\theta\\
\sin\theta & \cos\theta\\
\end{pmatrix}
\begin{pmatrix}
\nu_e\\
\nu_\mu\\
\end{pmatrix}
.
\end{eqnarray}
Using (\ref{eqn39}), the neutrino state (\ref{eqn37}) can be written in terms of flavor eigenstates as
\begin{eqnarray}\label{eqn40}
\begin{split}
&\ket{\upnu(r,t)} = \bigg[\cos^2\theta~\mathcal{F}_1(r,t)\Psi_0 + \sin^2\theta~\mathcal{F}_2(r,t)\Psi_0\bigg] \ket{\nu_e} \\&~~~~~~~~~~~ +  \bigg[ -\mathcal{F}_1(r,t)\Psi_0 + \mathcal{F}_2(r,t)\Psi_0 \bigg] \sin\theta \cos\theta \ket{\nu_\mu}.
\end{split}
\end{eqnarray}
The transition probability ratio is given by
\begin{eqnarray}\label{eqn41}
P_{e\mu} :=\frac{|\braket{\nu_\mu|\upnu(r,t)}|^2}{|\braket{\nu_e|\upnu(r_0,t_0)}|^2} =\frac{|\braket{\nu_\mu|\upnu(r,t)}|^2}{|\Psi_0|^2},
\end{eqnarray}
and by substituting from Eq.(\ref{eqn40}) we have
\begin{eqnarray}\label{eqn42}
\begin{split}
&P_{e\mu} = \frac{1}{4} \sin^2 (2 \theta) \bigg[\big|\mathcal{F}_1(r,t)\big|^2 + \big|\mathcal{F}_2(r,t)\big|^2 - 2\Re \{\mathcal{F}_1(r,t) \mathcal{F}^*_2(r,t) \} \bigg] \\& ~~~~~= \frac{1}{4} \sin^2 (2 \theta) \big[\mathcal{D}_{11} + \mathcal{D}_{22} - 2 \mathcal{D}_{12} \cos \Phi_{12} \big],
\end{split}
\end{eqnarray}
where $\Phi_{12}(r) = \varphi_1 - \varphi_2$ is the phase difference between different neutrino mass eigenstates. On the other hand, the survival probability ratio can be obtained by multiplying (\ref{eqn40}) by $\bra{\nu_e}$ from left-hand-side, so we have
\begin{eqnarray}\label{eqn43}
P_{ee} = \frac{|\braket{\nu_e|\upnu(r,t)}|^2}{|\Psi_0|^2},
\end{eqnarray}
and doing some manipulation leads to
\begin{eqnarray}\label{eqn44}
\begin{split}
&P_{ee} = \cos^4\theta \big|\mathcal{F}_1(r,t) \big|^2 + \sin^4\theta \big|\mathcal{F}_2(r,t) \big|^2 + \frac{1}{2} \sin^2 (2\theta) ~\Re\{\mathcal{F}_1(r,t) \mathcal{F}^*_2(r,t) \} \\& ~~~~ = \mathcal{D}_{11} \cos^4\theta + \mathcal{D}_{22} \sin^4\theta  + \frac{1}{2} \mathcal{D}_{12} \sin^2 (2\theta) \cos(\Phi_{12}).
\end{split}
\end{eqnarray}
These results are similar to the results obtained by Refs.\cite{Barger:1998xk,Blennow:2005yk}. The conformal coupling beside the influence on the oscillation phase through $\Phi_{12}$ has a damping or enhancing effect via  $\mathcal{D}_{ij}$:
\begin{equation}\label{sum}
P_{ee}+P_{e\mu}= \cos^2(\theta) \mathcal{D}_{11}+ \sin^2(\theta)\mathcal{D}_{22},
\end{equation}
which is not generally equal to one. For $\mathcal{D}_{ij}=1$, we have only oscillatory wavefunctions and the sum (\ref{sum}) equals one. Besides the neutrino flavor oscillation we may have neutrinos production or annihilation via the scalar field interaction. Note that if the conformal coupling is the same for all the neutrinos, we have $\mathcal{D}_{ij}=\frac{A_0^3}{A^3}$ for all $i, j$'s.

Let us give some examples: For the chameleon model, we have $A(\phi) = \exp(\beta\phi(r)/M_p)$ (see the appendix)
therefore
\begin{eqnarray}\label{eqn45}
\begin{split}
\mathcal{D}_{ij} = e^{-\frac{3}{2 M_p} [\beta_i\left(\phi_i(r) - \phi_i(r_0)\right) + \beta_j\left(\phi_j(r) - \phi_j(r_0)\right)]}.
\end{split}
\end{eqnarray}
Hence
\begin{eqnarray}\label{eqn46}
\begin{split}
&P_{ee} = \bigg[ e^{-\frac{3\beta_1}{2M_p}[\phi_1(r)-\phi_1(r_0)]} \cos^2\theta + e^{-\frac{3\beta_2}{2M_p}[\phi_2(r)-\phi_2(r_0)]} \sin^2\theta\bigg]^2 \\& ~~~~~~ - e^{-\frac{3}{2} \Sigma_{12}(r)} \sin^2(2\theta) \sin^2\bigg(\frac{\Phi_{12}(r)}{2}\bigg),
\\& P_{e\mu}  = e^{-\frac{3}{2} \Sigma_{12}(r)} \sin^2(2\theta) \bigg[\sin^2\bigg(\frac{\Phi_{12}(r)}{2}\bigg) + \sinh^2\bigg(\frac{3}{4} \Delta_{12}(r)\bigg)\bigg],
\end{split}
\end{eqnarray}
where $\Sigma_{ij}(r)\equiv  \frac{\beta_i [\phi_i(r) - \phi_i(r_0)]}{M_p} +  \frac{\beta_j [\phi_j(r) - \phi_j(r_0)]}{M_p}$ and $\Delta_{ij}(r) \equiv  \frac{\beta_i [\phi_i(r) - \phi_i(r_0)]}{M_p} - \frac{\beta_j [\phi_j(r) - \phi_j(r_0)]}{M_p}$.

If both mass eigenstates are coupled with a same coupling to the scalar field $\phi$, i.e.  $\beta_1 = \beta_2 = \beta$, the probabilities (\ref{eqn46}) will reduce to
\begin{eqnarray}\label{eqn47}
\begin{split}
& P_{ee} = e^{-\frac{3\beta}{M_p} [\phi(r) - \phi(r_0)]} \bigg[1 - \sin^2(2\theta) \sin^2\bigg(\frac{\Phi_{12}(r)}{2}\bigg)\bigg],
\\& P_{e\mu} = e^{-\frac{3\beta}{M_p} [\phi(r) - \phi(r_0)]} \sin^2(2\theta) \sin^2\bigg(\frac{\Phi_{12}(r)}{2}\bigg).
\end{split}
\end{eqnarray}
To obtain the above probabilities' values, we can use the solution to the chameleon equation of motion, which is solved numerically in subsection \ref{subapp1}. According to relations (\ref{eqn47}), not only the scalar field changes the oscillation phase but multiplies the probabilities by a scalar field dependent coefficient (for $\beta_1\neq \beta_2$ this coefficient becomes relevant for the relative number of flavors to each other). The sum of the probabilities is not one unless $\phi(r)=\phi(r_0)$. Hence generally
\begin{eqnarray}\label{eqn48}
P_{\textsf{tot.}} = e^{-\frac{3\beta}{M_p} [\phi(r) - \phi(r_0)]} \neq 1 ,
\end{eqnarray}
implying neutrino-scalar interaction, affecting the neutrino density.

Similarly for the symmetron model, the factor $\mathcal{D}$ is
\begin{eqnarray}\label{eqn49}
\mathcal{D} = [A A_{0}^{-1}]^{-3} \simeq 1 - \frac{3\left(\phi^2(r) - \phi^2(r_0)\right)}{2M^2}.
\end{eqnarray}
Using these relations leads us to the following probabilities
\begin{eqnarray}\label{eqn50}
\begin{split}
&P_{ee} = \bigg[1 - \frac{3\left(\phi^2(r) - \phi^2(r_0)\right)}{2M^2} \bigg]\bigg[1 - \sin^2(2\theta) \sin^2 \bigg(\frac{\Phi_{12}}{2}\bigg)\bigg],
\\& P_{e\mu} = \bigg[1 - \frac{3 \left(\phi^2(r) - \phi^2(r_0)\right)}{2M^2} \bigg] \sin^2(2\theta) \sin^2\bigg(\frac{\Phi_{12}}{2}\bigg).
\end{split}
\end{eqnarray}
We can also obtain the sum of probabilities above, which is independent of mixing angle $\theta$ and the phase shift $\Phi_{12}$,
\begin{eqnarray}\label{eqn51}
P_{\textsf{tot.}} = 1 - \frac{3 \left(\phi^2(r) - \phi^2(r_0)\right)}{2M^2} \ne 1.
\end{eqnarray}
Provided that the field $\phi(r)>\phi(r_0)$, the sum of the probabilities is smaller than unity and
$\mathcal{D}$ may be interpreted as a damping factor and conversely, for $\phi(r)<\phi(r_0)$ the neutrino density is enhanced.

\subsection{Three-flavor oscillations}\label{sub23}

We extend our results by taking three flavors into the account. As before, we first have to write the most general state of the neutrinos. The evolved neutrino state corresponding to the flavor $\alpha$ can be suggested as follows
\begin{eqnarray}\label{eqn55}
\ket{\upnu(r,t)}_\alpha = \sum_i \Psi_i(r_0,t_0)\mathcal{F}_i (r,t) U_{\alpha i}^* \ket{\nu_i},
\end{eqnarray}
where $\mathcal{F}_i(r,t)$ is a function which consists of $\mathcal{D}$ and phase factors. Note that $\mathcal{F}_i(r_0,t_0) = 1$, which explicitly implies that the initial condition can be satisfied by (\ref{eqn55}).
\\In order to obtain the different probabilities for this case, we shall replace the mass eigenstates by flavor eigenstates. As the two-flavor case, the mixing between various flavors of neutrinos can be described by a unitary mixing matrix $U$ as follows
\begin{eqnarray}\label{eqn56}
\ket{\upnu_i} = \sum_{\alpha} U_{\alpha i} \ket{\upnu_\alpha},
\end{eqnarray}
or equivalently
\begin{eqnarray}\label{eqn57}
\ket{\upnu_\alpha} = \sum_{i} U^*_{\alpha i} \ket{\upnu_i}.
\end{eqnarray}
Using relations (\ref{eqn56}) and (\ref{eqn57}), the state (\ref{eqn55}) can be written as
\begin{eqnarray}\label{eqn58}
\ket{\upnu(r,t)}_\alpha = \sum_i \sum_{\beta} \Psi_i(r_0,t_0)\mathcal{F}_i (r,t) U_{\alpha i}^* U_{\beta i} \ket{\nu_\beta}.
\end{eqnarray}
Assume that at $(r_0,t_0)$ we have a specific flavor neutrino, e.g. $\nu_\alpha$, requires that $\Psi_i(r_0,t_0)=\Psi_0$ for all $i$'s. The probability amplitude ratio ($\alpha\to \beta$) is then
\begin{eqnarray}\label{eqn59}
P_{\alpha \beta}&=&\frac{\bigg|\braket{\nu_\beta | \upnu(r,t)}_\alpha\bigg|^2}{\bigg|\braket{\nu_\alpha | \upnu(r_0,t_0)}_\alpha\bigg|^2} \nonumber \\
&=&\bigg| \sum_i \mathcal{F}_i(r,t) U^*_{\alpha i} U_{\beta i} \bigg|^2\nonumber \\
&=&\sum_{i,j} \mathcal{F}_i(r,t) \mathcal{F}_j^*(r,t) U^*_{\alpha i} U_{\beta i} U_{\alpha j} U^*_{\beta j}.
\end{eqnarray}
Therefore, the probability formula is given by
\begin{eqnarray}\label{eqn60}
\begin{split}
&P_{\alpha \beta} = \bigg| \sum_i \mathcal{F}_i(r,t) U^*_{\alpha i} U_{\beta i} \bigg|^2 \\&~~~~~ = \sum_{i,j} \mathcal{F}_i(r,t) \mathcal{F}_j^*(r,t) U^*_{\alpha i} U_{\beta i} U_{\alpha j} U^*_{\beta j}.
\end{split}
\end{eqnarray}
Now, we can calculate this formula in more details by using the explicit forms of the functions $\mathcal{F}_i(r,t)$ Eq.(\ref{me}), thus we have
\begin{eqnarray}\label{eqn61}
P_{\alpha\beta} = \sum_{i,j} \mathcal{D}_{ij} \mathcal{U}_{ij}^{\alpha \beta} e^{-i\Phi_{ij}},
\end{eqnarray}
where $\mathcal{U}_{ij}^{\alpha \beta} \equiv U^*_{\alpha i} U_{\beta i} U_{\alpha j} U^*_{\beta j}$ and $\Phi_{ij}$ is the phase shift. From properties $\mathcal{D}_{ij}(r) = \mathcal{D}_{ji}(r)$, $\mathcal{U}_{ji}^{\alpha \beta} = \big[\mathcal{U}_{ij}^{\alpha \beta}\big]^*$, $\Re[\mathcal{U}_{ij}^{\alpha \beta}] = \Re[\mathcal{U}_{ji}^{\alpha \beta}]$ and by doing some manipulation, the relation (\ref{eqn61}) can be rewritten as
\begin{eqnarray}\label{eqn62}
\begin{split}
& P_{\alpha\beta} = \sum_{i=1}^3 \mathcal{D}_{ii} \mathcal{U}_{ii}^{\alpha \beta} + 2\sum_{1 \leq i<j \leq 3} \mathcal{D}_{ij} \Re [\mathcal{U}_{ij}^{\alpha\beta}] \\& ~~~~~ - 4 \sum_{1 \leq i<j \leq 3} \mathcal{D}_{ij} \Re[\mathcal{U}_{ij}^{\alpha\beta}] \sin^2 \bigg(\frac{\Phi_{ij}}{2}\bigg) \\& ~~~~~ - 2 \sum_{1 \leq i<j \leq 3} \mathcal{D}_{ij} \Im[\mathcal{U}_{ij}^{\alpha\beta}] \sin (\Phi_{ij}),
\end{split}
\end{eqnarray}
where the two first non-oscillatory terms belong to the damped(enhancing)neutrino mixing. From another standpoint, the last term includes imaginary sector of the mixing matrix, so this term might correspond to the CP-violation.

To calculate probabilities $P_{ee}$, $P_{e\mu}$, etc., we have to apply the general form of the mixing matrix \cite{GonzalezGarcia:2007ib}
\begin{eqnarray}\label{eqn63}
U =
\begin{pmatrix}
c_{12} c_{13} & s_{12} c_{13} & s_{13}e^{-i\delta}\\
-s_{12} c_{23} - c_{12} s_{23} s_{13} e^{i\delta} & c_{12} c_{23} - s_{12} s_{23} s_{13} e^{i\delta} & s_{23} c_{13}\\
s_{12} s_{23} - c_{12} c_{23} s_{13} e^{i\delta} & -c_{12} s_{23} -s_{12} c_{23} s_{13} e^{i\delta} & c_{23} c_{13} \\
\end{pmatrix}
,
\end{eqnarray}
where $c_{ij} \equiv \cos\theta_{ij}$, $s_{ij} \equiv \sin\theta_{ij}$ and $\delta$ is the CP-violating phase. As an example, $\nu_e$-survival probability is given by
\begin{eqnarray}\label{eqn64}
\begin{split}
& P_{ee} = c_{12}^4 c_{13}^4 e^{-\frac{3}{2}\Sigma_{11}(r)} + s_{12}^4 c_{13}^4 e^{-\frac{3}{2}\Sigma_{22}(r)} + s_{13}^4 e^{-\frac{3}{2}\Sigma_{33}(r)} \\& ~~~~~ + 2 c_{12}^2 s_{12}^2 c_{13}^4 \cos(\Phi_{12}) e^{-\frac{3}{2}\Sigma_{12}(r)} \\& ~~~~~ + 2 c_{12}^2 c_{13}^2 s_{13}^2 \cos(\Phi_{13}) e^{-\frac{3}{2} \Sigma_{13}(r)} \\& ~~~~~ + 2 s_{12}^2 c_{13}^2 s_{13}^2 \cos(\Phi_{23}) e^{-\frac{3}{2} \Sigma_{23}(r)},
\end{split}
\end{eqnarray}
where $\Sigma_{ij}(r)$ is defined below Eq.(\ref{eqn46}). As can be seen, the probabilities for the three-flavor case are so tedious to calculate directly for the matrix elements of (\ref{eqn63}) and they are beyond our scope in this paper. Instead, we are going to investigate a simpler case in the next subsection.

\subsubsection{Neutrino decay: stable $\nu_1$  and unstable $\nu_2$}\label{subsub231}
We assume that the lightest neutrino mass eigenstate, $\nu_1$, is stable during neutrino propagation, i.e. $\mathcal{D}_{11} = 1$, whereas decay of $\nu_2$ as well as mixing among three neutrino eigenstates have effect on the deficit in the solar neutrino flux \cite{Joshipura:2002fb}. This statement means that we have only one coupled mass eigenstate affected by the scalar field. In such a case, the weak equivalence principle (WEP) is violated in this level ($\beta_1 =0$, whereas $\beta_2 \ne 0$). Note that there is also the possibility that WEP is violated at the macroscopic level in the screening models. But in regions where the scalar field is highly screened, WEP violation may not be detected by local gravitational tests \cite{Burrage:2017qrf,WEPtest}.

For this case, we assume that the $U_{e3}$ element of the matrix (\ref{eqn63}) is approximately negligible. The reason is that experiments result in small values for mixing angle $\theta_{13}$, e.g. in the Daya Bay ($\sin^2 2\theta_{13} = 0.0856 \pm 0.0029$) \cite{Adey:2018zwh} and RENO ($\sin^2 2\theta_{13} = 0.0896 \pm 0.0048(\text{stat.}) \pm 0.0047 (\text{syst.})$) \cite{Bak:2018ydk}. Therefore, the PMNS mixing matrix (\ref{eqn63}) reduces to
\[
U \cong
\begin{pmatrix}
c_{12} & s_{12} & 0\\
-s_{12} c_{23} & c_{12} c_{23} & s_{23}\\
s_{12} s_{23} & -c_{12} s_{23} & c_{23}\\
\end{pmatrix}
.
\]
Assuming $\ket{\nu_e}$ as the initial state and using the unitarity of the above matrix, we calculate different probabilities from Eq.(\ref{eqn62}) as follows
\begin{eqnarray}\label{eqn65}
\begin{split}
& P_{ee} = c^4_{12} + s^4_{12} \mathcal{D} +\frac{1}{2} \sqrt{\mathcal{D}} \sin^2(2\theta_{12}) \cos\Phi_{12} \\&
P_{e\mu} = \frac{1}{4} \sin^2(2\theta_{12}) c^2_{23} [1 + \mathcal{D} - 2\sqrt{\mathcal{D}} \cos\Phi_{12}]\\&
P_{e\tau} = \frac{1}{4} \sin^2(2\theta_{12}) s^2_{23} [1 + \mathcal{D} - 2\sqrt{\mathcal{D}} \cos\Phi_{12}],
\end{split}
\end{eqnarray}
where $\mathcal{D} \equiv \mathcal{D}_{22}$ is the $\mathcal{D}$ factor corresponding to $\nu_2$-decay. The observed neutrino rates in the SNO experiment from the CC (charged current), NC (neutral current) and ES (elastic scattering) interactions can be combined to place constraints on the separate $\upvarphi(\nu_e)$ and $\upvarphi(\nu_\mu) + \upvarphi(\nu_\tau)$ fluxes, which are respectively related to $P_{ee}$ and $P_{e\mu} + P_{e\tau}$. Both of these probabilities and also sum of them, i.e. $P_{\textsf{tot.}} = c^2_{12} + s^2_{12} \mathcal{D}$, are independent from atmospheric mixing angle $\theta_{23}$. On the other hand, the deficit in the rate of the solar neutrinos in this model on the Earth, i.e. $\delta P_{\alpha \beta} \equiv P^{(\textsf{undamped})}_{\alpha \beta} - P^{(\textsf{damped})}_{\alpha \beta}$, can be obtained as
\begin{eqnarray}\label{eqn66}
\delta P_{eX} \equiv 1 - P_{\textsf{tot.}} = (1-\mathcal{D}) s^2_{12},
\end{eqnarray}
The subscript $X$  refers to the particle corresponding to the (dark energy) scalar field ($\phi$) to which neutrinos can decay. This decay is provided by the conformal coupling, which leads to neutrino-$\phi$ interaction as can be seen from the equation of motion $\left(i\tilde{\gamma}^\mu \tilde{D}_\mu-m\right)\psi(x)=0$ (see Eq.(\ref{eqn15})), where $\tilde{\gamma}^\mu \tilde{D}_\mu$ is given by Eq.(\ref{eqn13}).
As neutrino and the scalar field interact through the conformal coupling, their density continuity equations are modified. This has been used in \cite{sad1} to alleviate the coincidence problem and to explain the onset of the present acceleration of the Universe in the non-relativistic era of mass varying neutrinos \cite{MV11}.

\begin{figure}[H]
	\centering
	\includegraphics[scale=0.5]{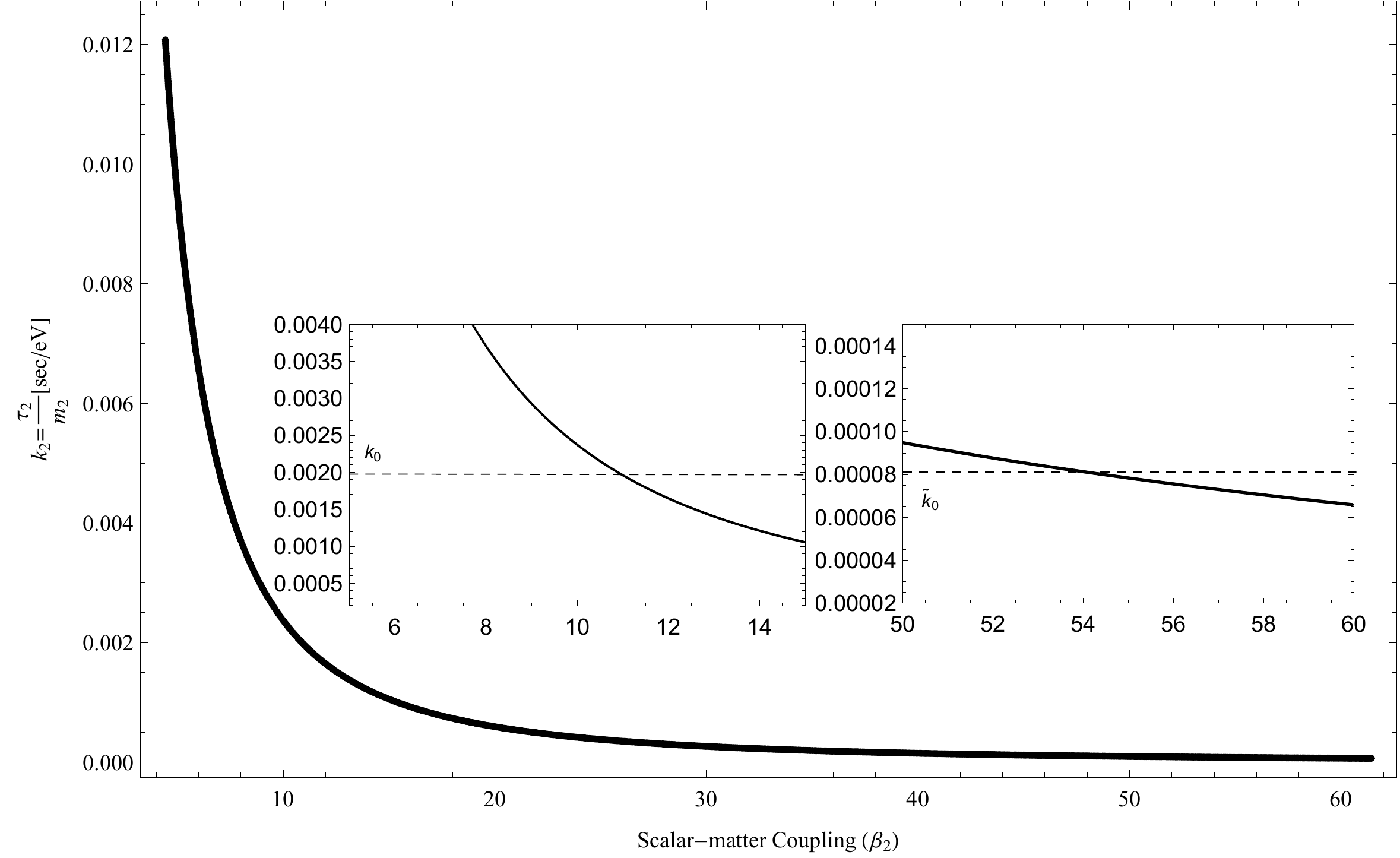}
	\caption{\footnotesize{Neutrino lifetime ratio $k_2$ versus $\beta_2$.}}
	\label{figk-Beta}
\end{figure}

As it can be seen from (\ref{eqn66}), $\delta P_{eX}$ depends on the mixing angle $\theta_{12}$ as well as the $\mathcal{D}$ factor. For increasing functions $\phi_i(r)$ we have $\mathcal{D} \leq 1$ and so there is a possibility for electron-neutrinos travelling from the source to the detector decaying to $\phi$ particles. Comparing the $\mathcal{D}$ factor of our model (Eq(\ref{eqn45}), with $\beta_1 =0$ and $\beta_2 \ne 0$) to the damping (or depletion) factor presented in \cite{Aharmim:2018fme,Joshipura:2002fb,Beacom:2002cb}, i.e. $\exp\left(-L/k E_\nu\right)$ suppresses the large values of the coupling parameter $\beta_2$. The solar neutrino lifetime $k_2 \left(=\tau_2/m_2\right)$ is then given by
\begin{eqnarray}\label{Comp}
k_2= \frac{2L M_p}{3\beta_2 E_\nu \left(\phi(L) - \phi(L_0)\right)},
\end{eqnarray}
where $L \simeq t$ (for ultra-relativistic neutrinos) is the distance between Earth and the Sun. The behavior of lifetime $k_2$ in terms of $\beta_2$ is plotted in Fig. \ref{figk-Beta}. We picked the numerical values $t \simeq 500$ sec. for distance from Earth to the Sun and neutrino energy $E_\nu \simeq 10$MeV for $^8$B solar neutrinos. Fitting all three phases of SNO data for $^8$B solar neutrinos \cite{Aharmim:2018fme,Aharmim:2011vm} and combined SNO + other solar neutrino experiments \cite{Aharmim:2018fme} yield $k_2 > \tilde{k}_0=8.08 \times 10^{-5}$ sec$/$eV at $90\%$ confidence ($\tilde{\beta}_2 < 54.1735$) and $k_2 > k_0 = 1.92 \times 10^{-3}$ sec$/$eV at $90\%$ confidence ($\beta_2 <11.1235$), respectively.
These requirements restrict the effective mass of the chameleon scalar field deep in space (in the dilute regions) (see (\ref{eqnA7})): $\tilde{m}_{\textsf{eff.}} < 2.99 \times 10^{-11}$eV (for $\tilde{\beta}_2$) and $m_{\textsf{eff.}} < 9.11 \times 10^{-12}$eV (for $\beta_2$).

For this case, we also obtain the discrepancy between undamped and damped probabilities for three cases separately. Using Eq.(\ref{eqn65}), we have
\begin{eqnarray}\label{eqn67}
\begin{split}
&\delta P_{ee} = s^2_{12} (1 - \sqrt{\mathcal{D}}) \big[(1 + \sqrt{\mathcal{D}}) s^2_{12} + 2 c^2_{12} \cos(\Phi_{12}) \big]\\&
\delta P_{e\mu} = c^2_{23} s^2_{12} (1 - \sqrt{\mathcal{D}}) \big[(1 + \sqrt{\mathcal{D}}) c^2_{12} - 2 c^2_{12} \cos(\Phi_{12}) \big]\\&
\delta P_{e\tau} = s^2_{23} s^2_{12} (1 - \sqrt{\mathcal{D}}) \big[(1 + \sqrt{\mathcal{D}}) c^2_{12} - 2 c^2_{12} \cos(\Phi_{12}) \big].
\end{split}
\end{eqnarray}
For example if we are interested in oscillation of neutrinos produced inside the Sun, for the case $\delta P_{ee} > 0$, we would have $\cos(\Phi_{12}) > - \tan^2\theta_{12}$, where we have assumed that $\sqrt{\mathcal{D}} \simeq 1$ outside the Sun, since both chameleon and symmetron fields are nearly constant deep in space. To have positive $\delta P_{e\mu}$ and $\delta P_{e\tau}$, however, the trivial condition $\cos(\Phi_{12}) <1$ has to be satisfied. Therefore, the resulting range for the phase difference is $-\tan^2\theta_{12} < \cos(\Phi_{12}) <1$.

\subsection{Flavor conversion in matter: The MSW effect}\label{sub24}
The weak interactions of neutrinos in matter modify the flavor conversion relative to the cases of propagation in the vacuum, as predicted decades ago by Mikheyev, Smirnov and Wolfenstein dubbed the MSW effect \cite{MSWWolf,MSWMS}.  After solar neutrinos production in the solar core and during their travels inside the Sun, they scatter forwardly from electrons until they leave the Sun and propagate through the vacuum to the Earth and finally detected, e.g. in the SNO  detectors. Therefore, we should take this effect into account. We start with the following rescaled Hamiltonian of the system
\begin{eqnarray}\label{eqn68}
\mathcal{H}^\prime = \frac{\Delta m^{\prime 2}}{4E}
\begin{bmatrix}
-\cos 2\theta + A & \sin 2\theta\\
\sin 2\theta & \cos 2\theta + A\\
\end{bmatrix}
,
\end{eqnarray}
where $A$ is a dimensionless parameter originated from $\nu_e - e^-$ scattering in matter, which is defined by
\begin{eqnarray}\label{eqn69}
A \equiv \frac{2\sqrt{2}G_F  E_\nu n_e}{\Delta m^{\prime 2}}
\end{eqnarray}
where $G_F$ is the weak Fermi constant and $n_e$ is the number density of electrons in the bulk of the Sun. Assuming $ A = \cos 2\theta$, leads us to the resonance point indicating maximal mixing in matter. The number density at the resonance point is then given by
\begin{eqnarray}\label{eqn70}
n_e(r_{\textsf{res}}) = \frac{\cos(2\theta) \Delta m^{\prime 2}(r_{\textsf{res}})}{2\sqrt{2}G_F E_\nu}.
\end{eqnarray}
According to Eq.(\ref{eqn69}) in the dilute regions, parameter $A$ takes a small value so that according to the Hamiltonian, Eq.(\ref{eqn68}), this implies an ordinary mixing. On the other hand, in the dense regions, $A$ is large enough to dominate in the diagonal elements of the Hamiltonian denoting mixing suppression in such regions, e.g. in the solar core, where neutrinos are born \cite{MohseniSadjadi:2017jne}.

Remember that the rescaled mass-squared splitting $\Delta m^{\prime 2}(r)$ depends explicitly on the scalar field values, see Eq.(\ref{eqn17}). The chameleon screening mechanism is specified by its varying mass so that in the dense regions, where parameter $n_e$ is considerable, this scalar field has a large mass hiding it from detectors. From properties of the chameleon (see the appendix), we see that chameleon takes small values $\phi_{\textsf{min},\textsf{in}} \ll M_p$ in such regions and, consequently, $\Delta m^{\prime 2} \longrightarrow \Delta m^2$. In the dilute regions, however, where the chameleon has a small mass and the parameter $A$ becomes small, the chameleon acquires larger values $\phi_{\textsf{min},\textsf{out}}\sim M_p$, which results in $\Delta m^{\prime 2} = \exp(2\beta\phi/M_p) \Delta m^2$.

Another screening mechanism, symmetron model,  is based on spontaneously symmetry breaking discussed in Appendix \ref{subapp2}. The $\mathbb{Z}_2$-symmetry will be broken in the low density environments, which yields $\phi_{\textsf{min},\textsf{out}} \ne 0$, see Eq.(\ref{eqnA15}), and parameter $A$ is small, so neutrinos experience ordinary oscillations. The restoration of the $\mathbb{Z}_2$-symmetry, on the other side, imposes $\phi_{\textsf{min},\textsf{in}} \longrightarrow 0$. In such a region, $A$ becomes dominant and, consequently, gives mixing suppression \cite{MohseniSadjadi:2017jne}.

The MSW flavor conversion in the Sun can be considered as a level crossing (or resonance) at which the most flavor change occurs when neutrinos cross this point. The first analytical formulas for adiabaticity violation in the Sun were calculated by Parke \cite{Parke:1986jy}. The standard expression for \textquotedblleft jumping\textquotedblright probability between $\nu_1$ and $\nu_2$ inside the Sun and at the resonance point is approximately given by \cite{Petcov:1987zj,Joshipura:2002fb,Bandyopadhyay:2001ct,Strumia:2006db}
\begin{eqnarray}\label{eqn71}
P_\mathcal{J} = \frac{e^{-\alpha \sin^2\theta} -e^{-\alpha}}{1-e^{-\alpha}}.
\end{eqnarray}
The parameter $\alpha$ in the exponent is equal to $\frac{\pi \Delta m^{\prime 2}}{E_\nu} \big|\frac{d\ln(n_e)}{dr}\big|_{r_{res.}}^{-1}$, where the electron number density of the Sun is considered to have approximately an exponential form, i.e. $\propto \exp(-r/R_{\odot})$ \cite{Strumia:2006db}.
As in Ref.\cite{Bandyopadhyay:2001ct}, we prefer to use the LMA (large mixing angle) solutions when we consider neutrino decay as well as neutrino mixing. The observation of electron antineutrino ($\bar{\nu}_e$) flavor oscillations by KamLAND \cite{Gando:2013nba} presented that neutrino mixing is mainly responsible for what had been known as solar neutrino problem. The mixing angle ($\theta_{12}$) extracted by KamLAND \cite{Gando:2013nba} is quite consistent with MSW-LMA solution obtained by solar neutrino experiments such as SNO \cite{Aharmim:2011vm} and SK \cite{SK}. The coincidence of the mixing parameters ($\theta_{12}$, $\Delta m^2_{21}$) determined in KamLAND and solar neutrino experiments may imply the confirmation of CPT invariance \cite{Smirnov:2016xzf}.

To discuss the probabilities from the MSW effect \cite{Bandyopadhyay:2001ct,Joshipura:2002fb} in the Sun, which are a mixture of flavor oscillation and neutrino decay, we proceed as follows. As discussed in subsection \ref{subsub231}, $\nu_2$-instability with mixing together cause the solar neutrino problem, whereas the mass eigenstate $\nu_1$  remains stable, thus, $P_1$ and $P_2 . \mathcal{D}$ are respectively the probabilities of detecting $\nu_1$ and $\nu_2$ on the Earth. $P_1$ ($P_2$) defined as the probability of $\nu_e \longrightarrow \nu_1$ ($\nu_2$) can be written as
\begin{eqnarray}\label{eqn72}
P_1 = 1- P_2 = P_{\mathcal{J}} \sin^2 \theta_{\textsf{m}} + (1-P_{\mathcal{J}}) \cos^2 \theta_{\textsf{m}},
\end{eqnarray}
where $\theta_{\textsf{m}}$ is the mixing angle in the matter defined by $\tan 2\theta_{\textsf{m}} = \frac{\sin 2\theta}{\cos 2\theta-A}$ \cite{Guidry:2018ocm}. Then using the unitary mixing matrix in the subsection \ref{subsub231}, the probability formulas are given by
\begin{eqnarray}\label{eqn73}
\begin{split}
& P_{ee} = c^2_{12} P_1 + \mathcal{D}^{\odot} s^2_{12} P_2 \\&
P_{e\mu} = c^2_{23} s^2_{12} P_1 + \mathcal{D}^{\odot} c^2_{23} c^2_{12} P_2 \\&
P_{e\tau} = s^2_{23} s^2_{12} P_1 + \mathcal{D}^{\odot} s^2_{23} c^2_{12} P_2,
\end{split}
\end{eqnarray}
where $\mathcal{D}^\odot$ is the $\mathcal{D}$ factor inside the Sun and we have also considered $\mathcal{D}^{\textsf{vac}} \simeq 1$. To see the damping (decay) behavior only, the phase parts in the probabilities have been ignored.

\section{Results, discussions and conclusion}\label{sec3}

We have studied a scenario about damped neutrino oscillations in a curved spacetime, which consists of a scalar field conformally coupled to other ingredients such as matter and neutrinos (see Eq.(\ref{eqn2})).

To derive the oscillation probabilities, we studied the behavior of the Dirac equation under the conformal transformation, which reduces the model into the flat spacetime with a rescaled wavefunction and a coordinate dependent mass (see Eqs.(\ref{eqn17}) and (\ref{eqn18})). By solving the mass-varying Dirac equation in the flat spacetime, we derived mass-eigenstates of the neutrinos (see Eq. (\ref{me})). The presence of factor $\mathcal{D}_{ij}$ (see (\ref{eqn35})) and the deficiency in the total probability can be interpreted as the interaction of the neutrino with the scalar field, allowing them to convert to each other.
To be more specific, we considered solar neutrinos and consider two examples: the chameleon and the symmetron models, which we review briefly and point out the required relations for our discussion in the appendix.  In these models, the effective masses depend explicitly on the local matter density, see Eqs.(\ref{eqnA7}) and (\ref{eqnA17}), leading to the screening effect in a dense area.

The dynamics of the scalar field is determined with a good approximation by the matter density. We first expand the scalar field around its background value and obtain an equation for the fluctuation(see (\ref{eqnA9})). Inside the Sun, the density ( $\rho_{\odot}(R)$) is shown in \ref{fig8}, and the scalar field equation is numerically depicted in Figs.\ref{fig9} and \ref{fig10} for various values of the coupling parameter $\beta$ for the chameleon, and is depicted in \ref{fig11} for the symmetron.

As can be seen from (\ref{eqn36}), the chameleon scalar field affects the oscillation phase. We then found the formulas for damped transition and survival probabilities and also obtained a value for violation in the total probability conservation (see Eq.(\ref{eqn66})) for both two-flavor and three-flavor neutrino oscillations. Finally, we studied the non-oscillatory effects of neutrino forward elastic scattering from electrons in matter, MSW effect, in subsection \ref{sub24}. As we concluded, the LMA solutions of neutrino oscillations are the best solutions describing the decay process in this model.

\begin{figure}[H]
\begin{subfigure}{.5\textwidth}
\centering
\includegraphics[scale=0.46]{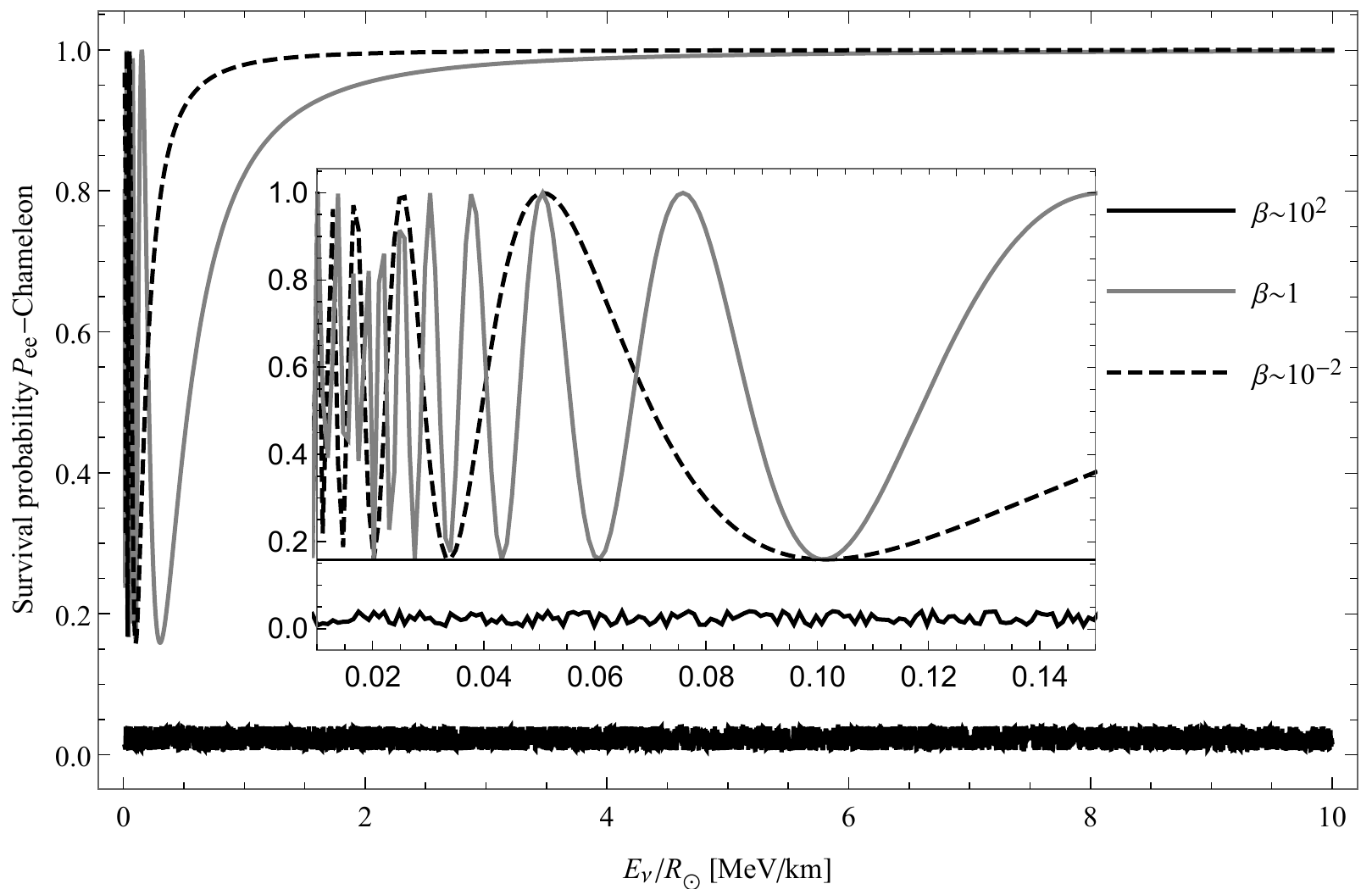}
\caption{}
\label{fig2leftup}
\end{subfigure}
\begin{subfigure}{.5\textwidth}
\centering
\includegraphics[scale=0.45]{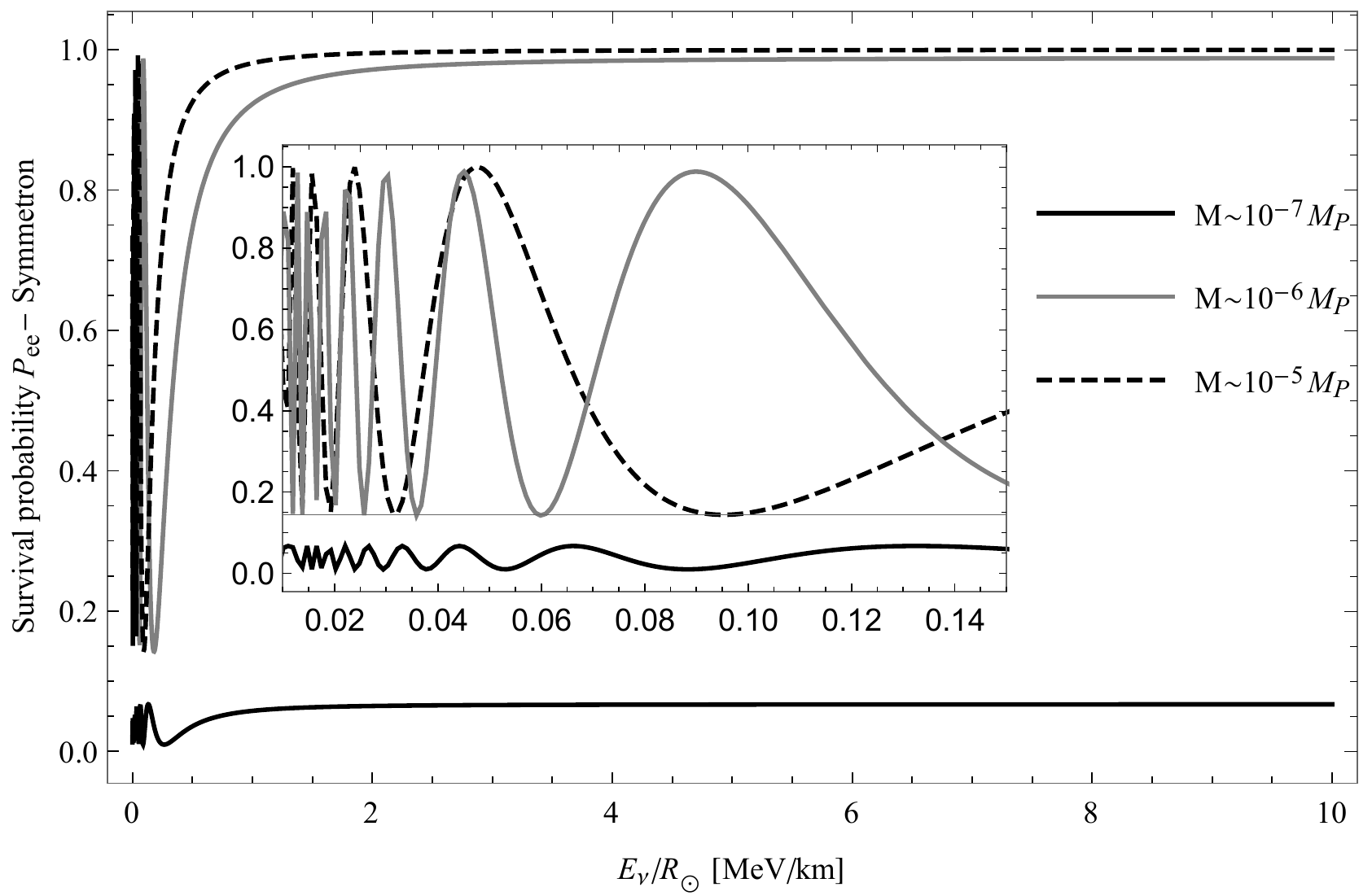}
\caption{}
\label{fig2rightup}
\end{subfigure}
\caption{\footnotesize{Plots of $P_{ee}$ in the presence of chameleon and symmetron scalar fields for the two-flavor neutrino oscillations. Both panels are figures depicted for the survival probability in terms of (neutrino energy/solar radius). Different curves describe $P_{ee}$ corresponding to various values of the chameleon-matter coupling $\beta$, see left plot. Right panel, however, is plotted for $P_{ee}$ in the presence of symmetron for $\mu \sim 10^{-12}$eV, $\lambda \sim 10^{-50}$ and for three various values of mass-scale parameter $M$. The solid horizontal line in each inset shows the minimum values.}}
\label{fig2}
\end{figure}

Now let us illustrate our results via some numerical examples.
In Fig.\ref{fig2}, we plot the survival probability $P_{ee}$ as a function of (neutrino energy/solar radius) for the two-flavor case to see the effects of the conformal coupling on the oscillation phase. It should be noticed that we have picked the numerical values $n=1$, $M \simeq 2.08$keV \cite{Waterhouse:2006wv}, the mass-squared splitting $\Delta m^2 = 7.4 \times 10^{-5}$eV$^2$ and the mixing angle $\tan^2\theta = 0.41$ for the LMA solution of neutrino oscillations \cite{Esteban:2020cvm}. We have also picked three values for coupling strength $\beta$ in the chameleon model, panel \ref{fig2leftup}. In the chameleon model proposed by Khoury and Weltman \cite{Khoury:2003aq}, scalar field is coupled to matter with gravitational-strength, i.e. $\beta \sim \mathcal{O}(1)$ (gray curve) in agreement with expectations of the string theory \cite{Mota:2006fz}, or smaller, i.e. $\beta \ll 1$. Setting $\beta \sim 0$ reproduces the no-chameleon figure (black-dashed curve). On the other hand, as the satellite experiments \cite{Touboul:2008zz,Mester:2001,Nobili:2000} have proposed, they are unable to put an upper bound on $\beta$ \cite{Mota:2006fz}, so we plotted a figure for $\beta \sim \mathcal{O}(10^2)$ (black-solid curve). The resulting profile for the survival probability is sensitive to $\beta$ such that this probability will be suppressed with intense oscillations when $\beta$ increases. Note that the choice $\beta = \mathcal{O}(10^2)$ is not suitable as it gives $P_{ee} =0.019$ at energy $E_\nu =10$MeV, which is negligible comparing to the experimental values ($P^{\textsf{SNO}}_{ee} = 0.340 \pm 0.023$ \cite{Sun2} and $P^{\textsf{Borexino}}_{ee} = 0.350 \pm 0.090$ \cite{Agostini:2018uly} for $^8$B neutrinos).  Also, the probability $P_{ee}$ and its oscillating behavior affected by the symmetron mechanism is presented in the panel \ref{fig2rightup} for three values of mass-scale parameter $M \lesssim 10^{-4} M_p$. This constraint on $M$ is imposed by local tests of gravity \cite{Hinterbichler:2011ca,Hinterbichler:2010es}. Note that $\mu \sim 10^{-12}$eV \cite{MohseniSadjadi:2017jne}, $\lambda \sim 10^{-50}$ ($\lambda \gtrsim 10^{-96}$ \cite{Hinterbichler:2011ca}). As shown in this figure, the choice $M \lesssim 10^{-7} M_p$ is not consistent with observational results.

\begin{figure}[H]
\centering
\includegraphics[scale=0.4]{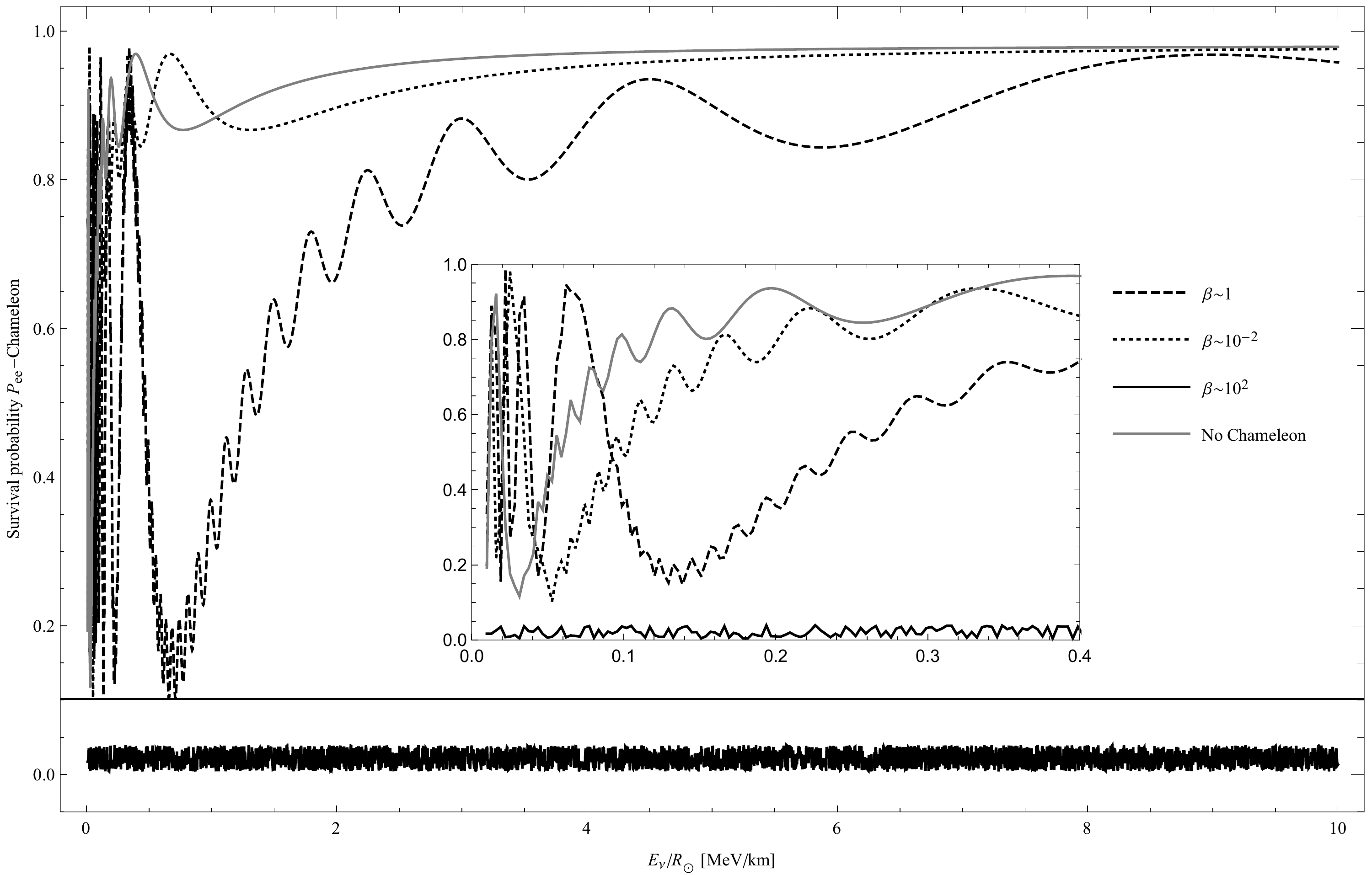}
\caption{\footnotesize{In this figure, we have depicted the electron-neutrino survival probability in terms of ratio $E_{\nu}/R_{\odot}$, for $n=1$ and $M=2.08$keV. Different curves describe the $P_{ee}$ for various values of the coupling parameter $\beta$. Rapid oscillations by increasing $\beta$ is a result of the chameleon-dependent oscillation phase. Gray curve represents the survival probability in the absence of the corresponding scalar field. The solid horizontal line shows the minimum values.}}
\label{fig3}
\end{figure}

Fig.\ref{fig3} shows the survival probability as a function of ratio (neutrino energy$/$solar radius) assuming $\tan^2 \theta_{12}=0.41$, $\sin^2 2\theta_{23} \simeq 0.99$, $\sin^22\theta_{13} \simeq 0.09$ and
\[ \Delta m_{21}^2=7.4 \times 10^{-5} eV^2 \] and \[|\Delta m_{23}^2|=2.5 \times 10^{-3} eV^2\]
for the three-flavor case \cite{Esteban:2020cvm} in the presence of the chameleon scalar field. We have also chosen the numerical values $n=1$, $M=2.08$keV \cite{Waterhouse:2006wv}. $P_{ee}$ is depicted for various orders of magnitude of coupling parameter to compare the probability amplitude and phase of oscillations for different $\beta$'s. Gray curve is plotted ignoring the effects of the chameleon. As can be seen, the probability amplitude is suppressed with rapid oscillations for $\beta \gtrsim 10^2$ (black-solid curve) such that it has intersection with none of experimental values for survival probability of $^8$B neutrinos ($E_\nu \simeq 10 $ MeV) measured by SNO ($P^{\textsf{SNO}}_{ee} = 0.340 \pm 0.023$) \cite{Sun2} and Borexino ($P^{\textsf{Borexino}}_{ee} = 0.350 \pm 0.090$) \cite{Agostini:2018uly}, for instance. As an example, $P_{ee} = 0.349$ for $\beta \sim 1$ and energy $E_{\nu} = 10$MeV ($^8$B solar neutrinos), whereas $P_{ee} = 0.038$ for $\beta \sim 10^2$ and the same energy. This inconsistency in survival probability for large $\beta$ can be addressed to some similar discussions in section \ref{subsub231}.

\begin{figure}[H]
	\centering
	\includegraphics[scale=0.4]{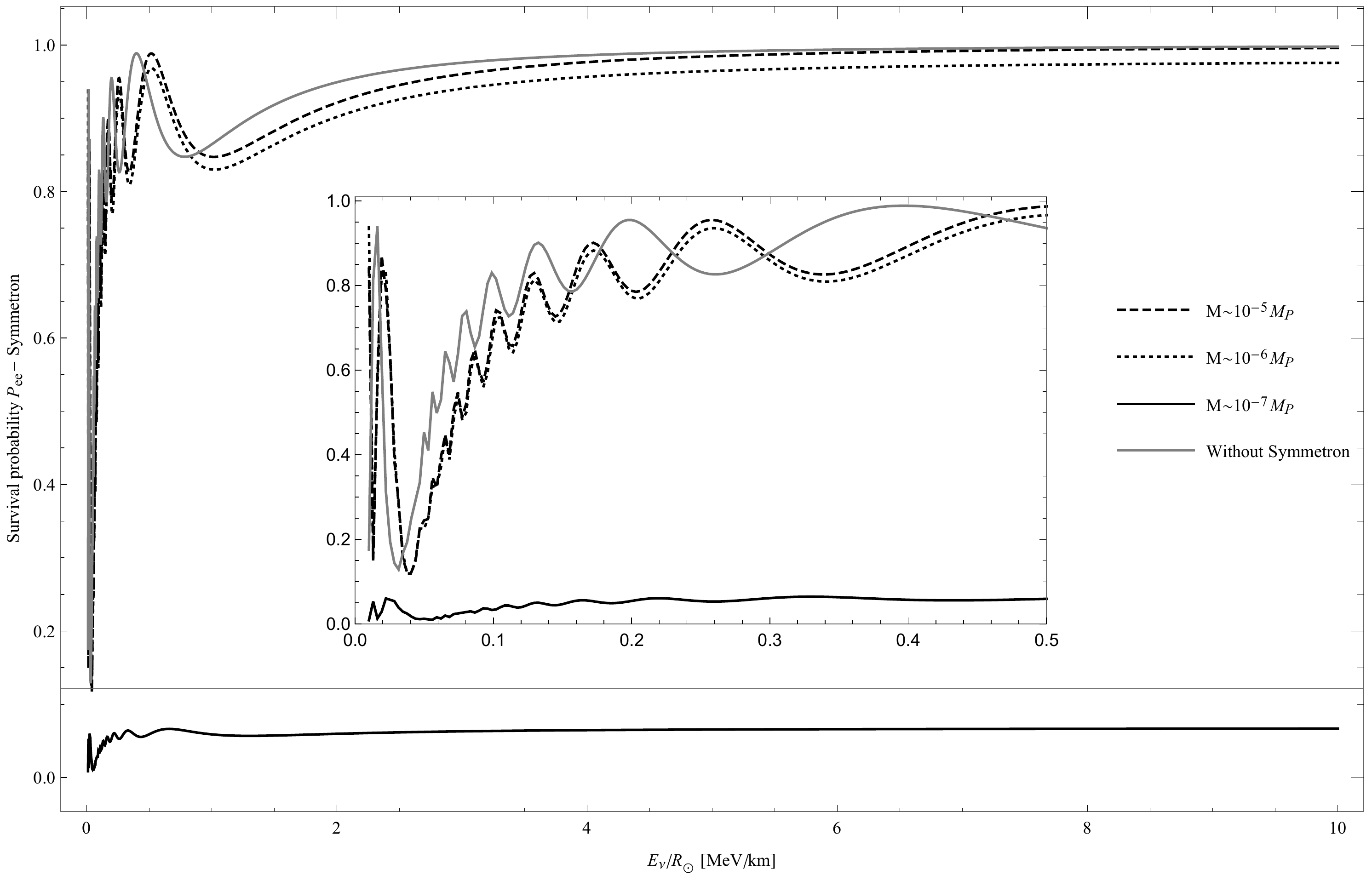}
	\caption{\footnotesize{This figure describes the behavior of the $P_{ee}$ in the presence of the symmetron scalar field, for $\mu \sim 10^{-12}$eV, $\lambda \sim 10^{-50}$ and for three various values of mass-scale parameter $M$. Gray curve represents the survival probability in the absence of the corresponding scalar field.}}
	\label{fig4}
\end{figure}

 Figure \ref{fig4}, however, shows the effects of the symmetron on the $P_{ee}$-amplitude and its phase. Different values of $M \lesssim 10^{-4} M_{p}$ and also a case in the absence of symmetron (gray curve) are considered. As a numerical example, survival probability is equal to $P_{ee} = 0.067$ for $M=10^{-7}M_p$ and energy $E_{\nu}=10$MeV. As a result, probability $P_{ee}$ will be suppressed for $M \lesssim 10^{-7} M_{p}$, with no experimental evidence (see black-solid curve) \cite{Sun2,Agostini:2018uly}.

\begin{figure}[H]
	\centering
	\includegraphics[scale=1]{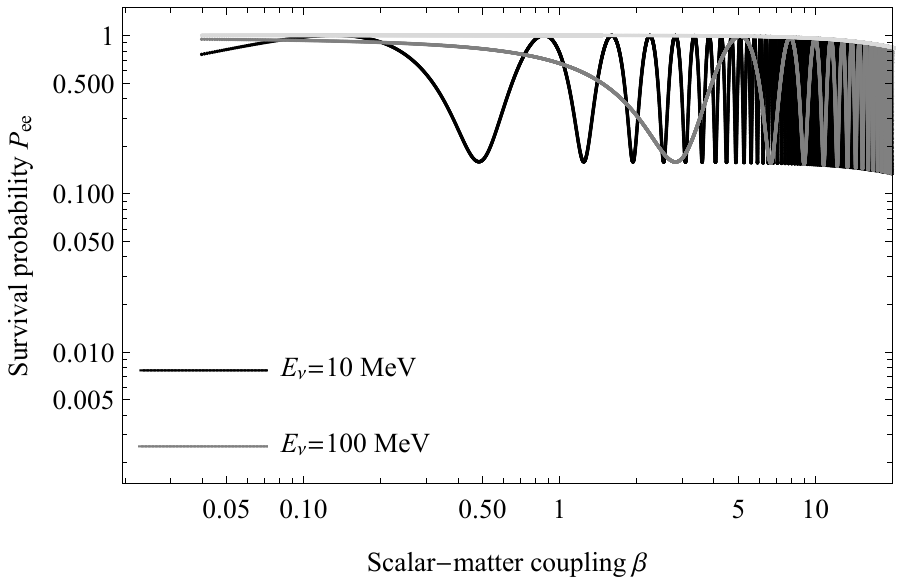}
	\caption{\footnotesize{The behavior of $P_{ee}$ versus the chameleon-matter coupling $10^{-2} \leq \beta \leq 15$ for solar and high energy neutrinos. Note that $\Delta m^2 = 7.4 \times 10^{-5}$eV$^2$ and $\tan^2 \theta = 0.41$. The envelope function (light-gray curve) shows how $P_{ee}$ gradually drops when $\beta$ increases, due to the existence of $\mathcal{D}$ factor.}}
	\label{fig5}
\end{figure}

The electron-neutrino survival probability behavior is also shown as a function of chameleon-matter coupling $\beta$ in Fig.\ref{fig5}.

Fig.\ref{fig5} shows a damped oscillation, which its damped part illustrated by the upper envelope curve (light-gray curve) is trivial because of the $\mathcal{D}$ factor behind the oscillatory term, which is responsible for its oscillation. As can be easily seen, $P_{ee}$ has rapid oscillations for increasing $\beta$, which is a clear sign of the effect of the chameleon on the phase. Note that we have used the LMA mass-squared splitting and mixing angle, as mentioned before, and this figure has been plotted for solar and high energy neutrinos as shown in the legends.

The solar neutrino problem can be solved by considering decay processes as well as mixing. There are generally two kinds of decays depending on whether the final state particles are only \textit{invisible}, such as $\phi$-particles, sterile neutrinos (generally non-active neutrino flavors), Majorons, etc., or they include \textit{visible} particles too, e.g., active neutrino flavors \cite{Baerwald:2012kc}. An applied fit to BS05(OP) data of solar neutrinos \cite{Bahcall:2004pz} can lead us to the zeroth-order measurements $0.292^{+0.067}_{-0.039}$ and $0.12^{+0.14}_{-0.23}$ for the electron-neutrino survival probability and the conversion probability to unknown states respectively \cite{Drouin:2011pna}. The latter value of the conversion probability (neutrino decay to unknown states) can be interpreted as $\phi$-particles.

\begin{figure}[H]
	\begin{subfigure}{.5\textwidth}
		\centering
		\includegraphics[scale=0.59]{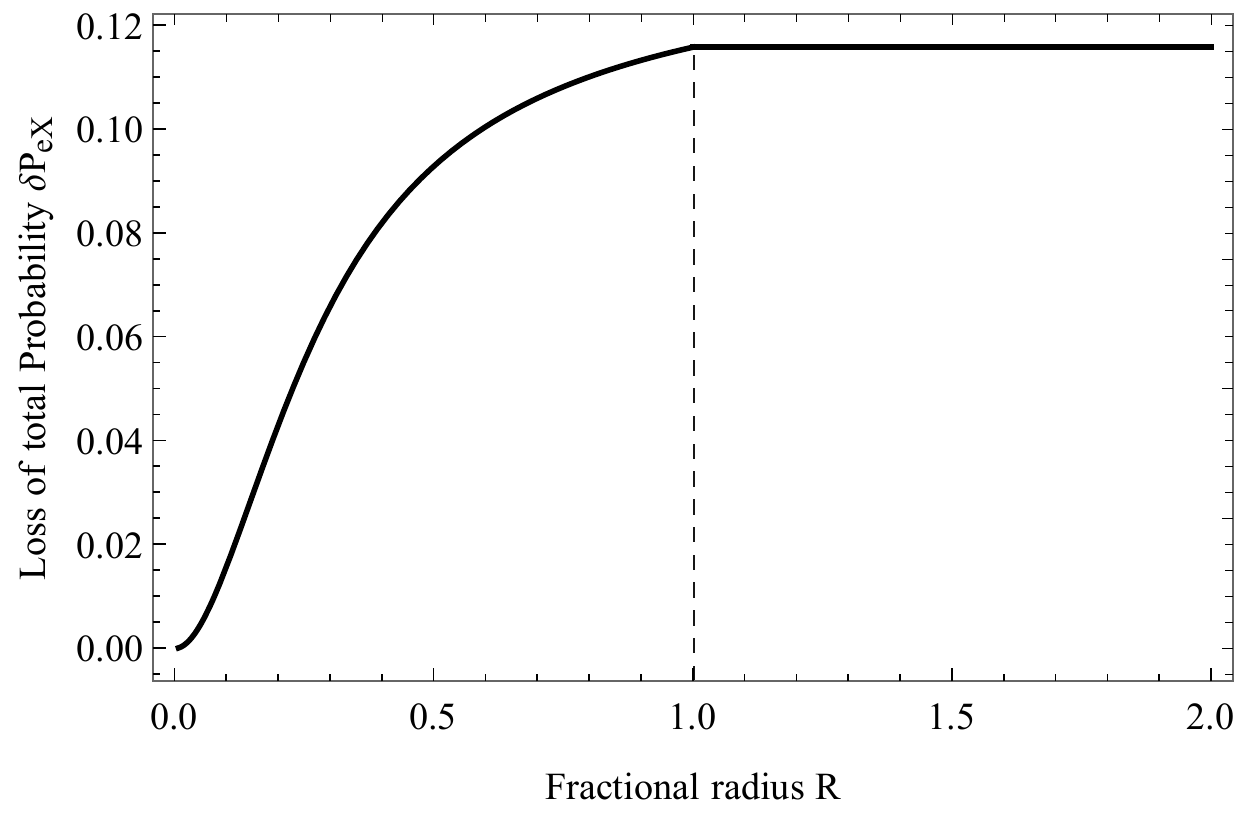}
		\caption{}
		\label{fig6left}
	\end{subfigure}
	\begin{subfigure}{.5\textwidth}
		\centering
		\includegraphics[scale=0.63]{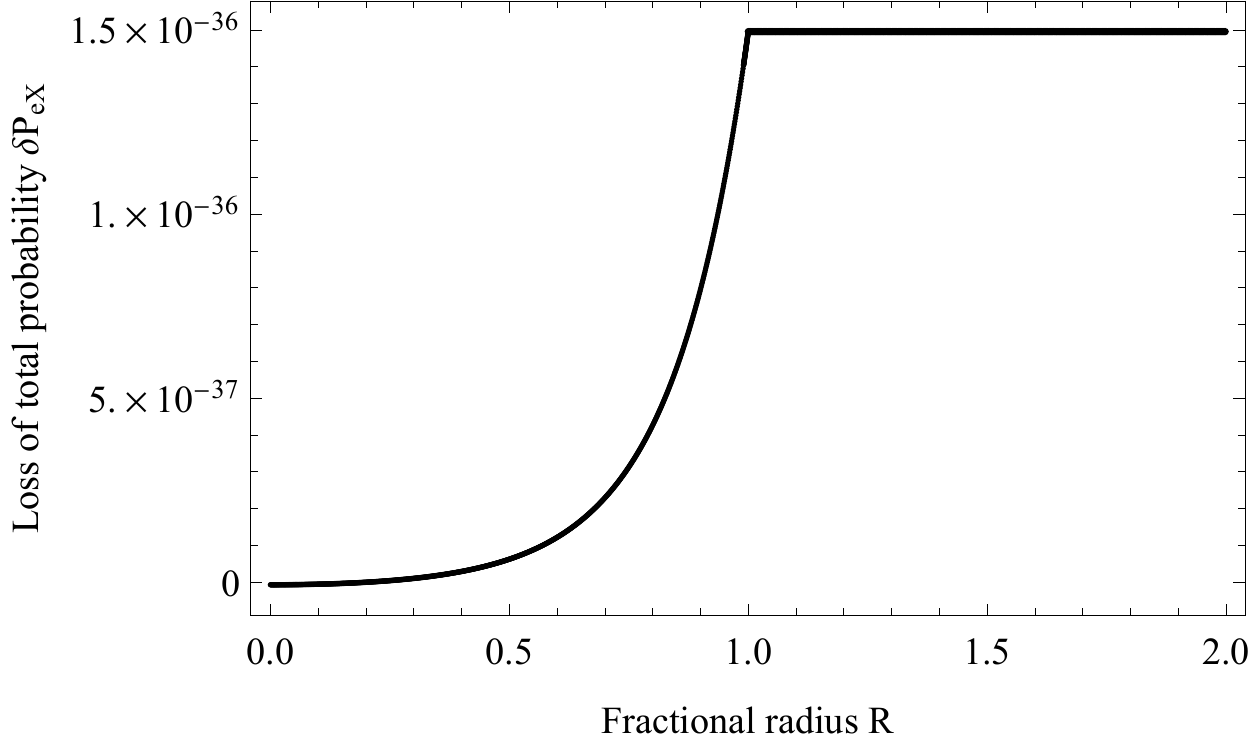}
		\caption{}
		\label{fig6right}
	\end{subfigure}
	\caption{\footnotesize{The losses in the total probability for the chameleon model with $\beta \sim 1$ (a) and for the symmetron model with $M \sim 10^{-4} M_{{p}}$ (b), inside and outside the Sun.}}
	\label{fig6}
\end{figure}

To clarify our model, we take a numerical example for $\delta P_{eX}$. We need first to look for the numerical values of the field, which minimize the effective potential for background matter density on the cosmological scales, i.e. $\rho_0 \sim 10^{-24}$ g.cm$^{-3}$ \cite{Spergel:2006hy}. This means that we first set $V_{\textsf{eff.},\phi} = 0$ (see Eq.(\ref{eqnA6})), then we have
\[
\frac{\phi_{0}}{M_p} = \bigg[ \frac{n M^{4+n}}{\beta \rho_0 M^n_p} \bigg]^{\frac{1}{n+1}}.
\]
Setting parameters $n = 1 $, $\beta \simeq 1$ and $ M = 2.08$ keV \cite{Waterhouse:2006wv} for the chameleon field gives us $\phi_0 \approx 0.609074 M_p$. As mentioned earlier, our choice of coupling parameter $\beta$ is consistent with gravitation-strength interactions \cite{Upadhye:2012qu}. Using all these and also the numerical values for the symmetron field, i.e. $\mu = 10^{-12}$, $\lambda = 10^{-50}$ and $M \simeq 10^{-4} M_p $, we plot Figure \ref{fig6}, which shows the amount of the discrepancy in the damped total probability from unity for two cases of the chameleon and symmetron scalar fields. In both panels, the discrepancy grows to a constant value at $R=1$ and remains constant outside the body to the Earth. The left panel of Fig.\ref{fig6} shows that the numerical behavior outside the Sun agrees well with the numerical value presented above. The discrepancy in total probability approaches to the value $\delta P_{eX} \simeq 0.118$. This value is much larger than that for the symmetron field depicted in the right panel of Fig.\ref{fig6}. This difference may refer to their different coupling functions.

\begin{figure}[H]
	\begin{subfigure}{.5\textwidth}
		\centering
		\includegraphics[scale=0.35]{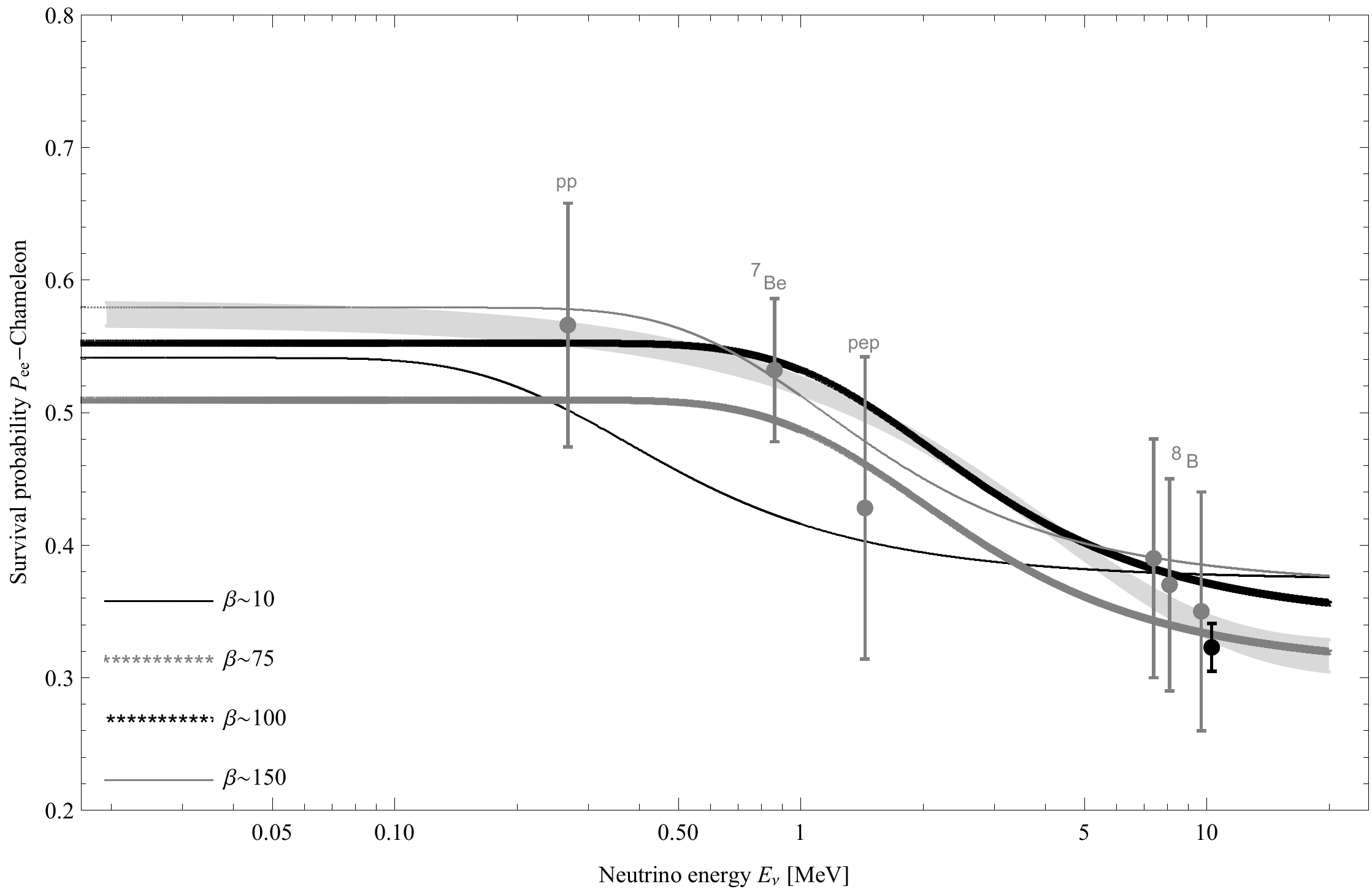}
		\caption{}
		\label{fig7left}
	\end{subfigure}
	\begin{subfigure}{.5\textwidth}
		\centering
		\includegraphics[scale=0.37]{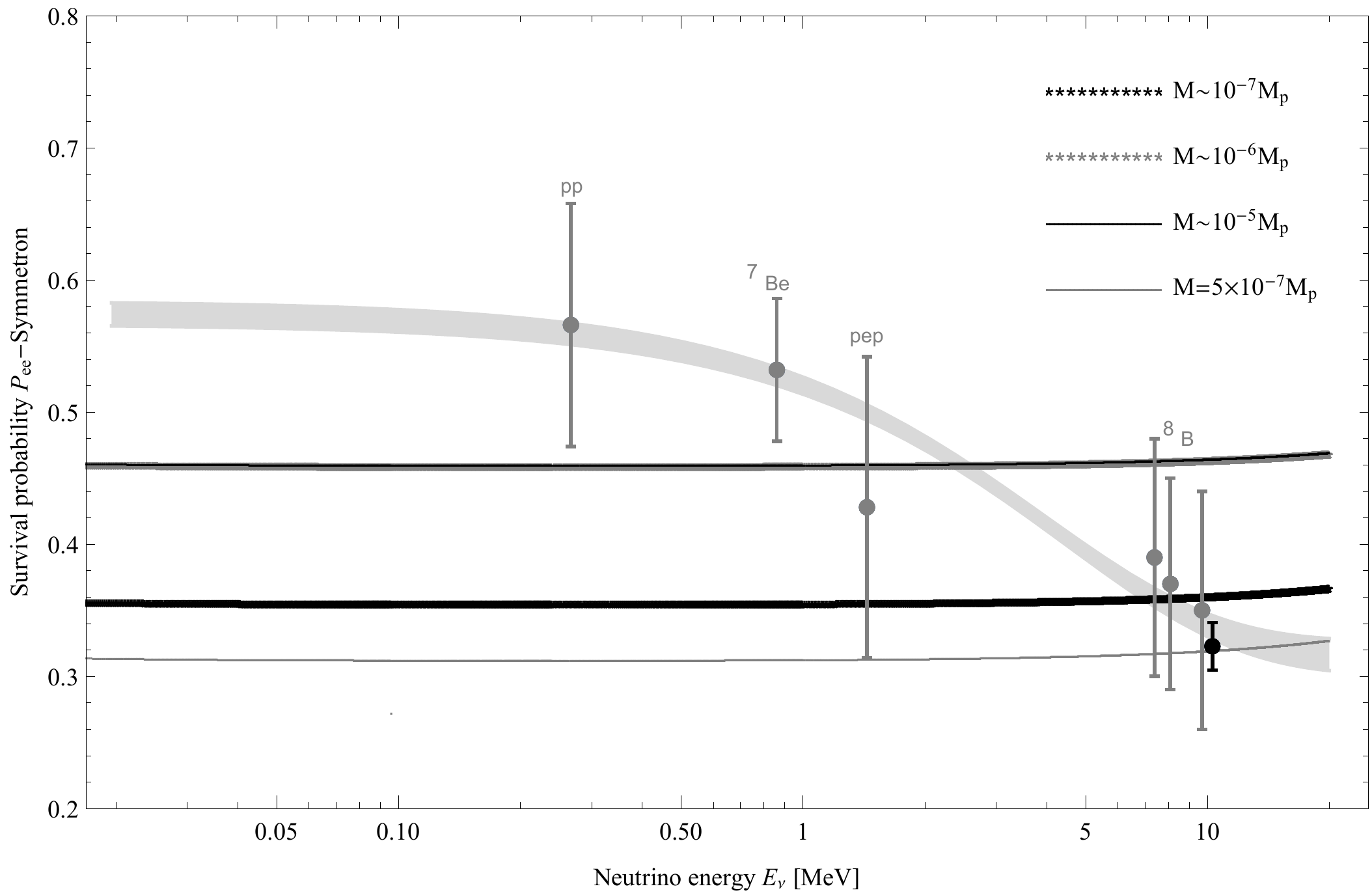}
		\caption{}
		\label{fig7right}
	\end{subfigure}
	\caption{\footnotesize{Electron-neutrino survival probability on the Earth as a function of its energy. The left panel describes $P_{ee}$ including the MSW-LMA effect in the presence of the chameleon scalar field for four different $\beta$'s, whereas the right panel shows how the MSW effect governs $P_{ee}$ in the presence of the symmetron scalar field for four different values of mass-scale $M$. In panel (b) curves of $M \sim 10^{-5}M_p$ and $M \sim 10^{-6} M_p$ are so close such that they seem overlapped. Note that the light-gray band in both panels illustrates the theoretical prediction of $\nu_e$ survival probability by Borexino for MSW-LMA within $\pm 1 \sigma$. Experimental values are also shown for Borexino (gray) and SNO + SK $^8$B (black) data.}}
	\label{fig7}
\end{figure}

Figure \ref{fig7} illustrates the effects of matter (in the bulk of the Sun) called MSW effect on solar neutrinos discussed in the subsection \ref{sub24}. Both plots show the electron-neutrino survival probability on the Earth as a function of its energy in MeV. We have also depicted the $P_{ee}$-experimental values from the Borexino data \cite{Agostini:2018uly} of pp, $^7$Be, pep and $^8$B fluxes (gray points). Black point is also represents the SNO + SK $^8$B data \cite{Zyla:2020zbs}. Light-gray band in both panels is the best theoretical prediction of $P_{ee}$ (within $\pm 1 \sigma$) according to MSW-LMA solution \cite{Agostini:2018uly}. We guess that the best fit to this curve can be written as
\[P_{ee}\left(E_{\nu}[MeV]\right) \simeq 0.322 + 0.244 e^{-0.25 \left(\frac{E_{\nu}}{MeV}\right)} .\]

To draw these figures, we have assumed that $\Delta m^{\prime 2} \simeq 2.49  \times 10^{-4}$eV$^2$, $\tan^2 \theta_{\textsf{m}} \simeq 0.43$ and $\sqrt{2} G_F n_e \simeq 2.27 \times 10^{-7} \left(\frac{eV^2}{MeV}\right)$ \cite{Guidry:2018ocm}. Different curves have been plotted according to the present model of MSW effect, which are approximately consistent with experimental (Borexino) values in low energy range (pp, $^7$Be and pep) and also (Borexino and SNO + SK) in high energy range ($^8$B), where changing flavor for the latter range is caused mostly by matter effects in the bulk of the Sun \cite{Agostini:2018uly}.

Data analysis of future neutrino flavor conversion measurements, as has already been done by neutrino experiments such as SK \cite{SK,SKnonstand}, HK \cite{HK} and JUNO \cite{JUNO}, determining the different mixing parameters in the framework of the nonstandard scalar-neutrino interactions might improve the values of these parameters. Besides, such nonstandard interaction effects on the mixing parameters will help explain the difference of $\Delta m_{21}^2$ extracted by solar neutrino and KamLAND experiments \cite{Smirnov:2016xzf,Liao:2017awz}.

Deep Underground Neutrino Experiment (DUNE) \cite{DUNE}, as a long-baseline neutrino experiment, with a well-understood beam and trajectory is ideal for probing matter-scalar field nonstandard interactions affected by the mass density of the Earth.
DUNE will collect much more data than the current experiments with improved systematic uncertainties \cite{Liao:2016orc,DUNEnonstand}, which might help to reach a higher sensitivity to the neutrino-scalar interactions and active-to-sterile neutrino mixing.
Since the chameleon and symmetron are environment-dependent scalar fields, the Earth's density \cite{ErthDen} might be taken into account in neutrino-scalar coupling discussion, as has been done for the Sun in the present paper. Furthermore, to restrict the coupling parameter $\beta_i$ more precisely by the constraints on neutrino lifetime, the DUNE data will be used \cite{Choubey:2017dyu}.

\appendix
\numberwithin{equation}{section}
\makeatletter
\newcommand{\section@cntformat}{Appendix \thesection:\ }
\makeatother

\section{Appendix: Screening mechanism}\label{app1}
This section provides a brief review of two screening models: the chameleon and the symmetron models, and their equations of motion in a nearly flat static spherically symmetric spacetime.
\subsection{Chameleon mechanism}\label{subapp1}
\begin{figure}[H]
	\centering
	\includegraphics[scale=0.7]{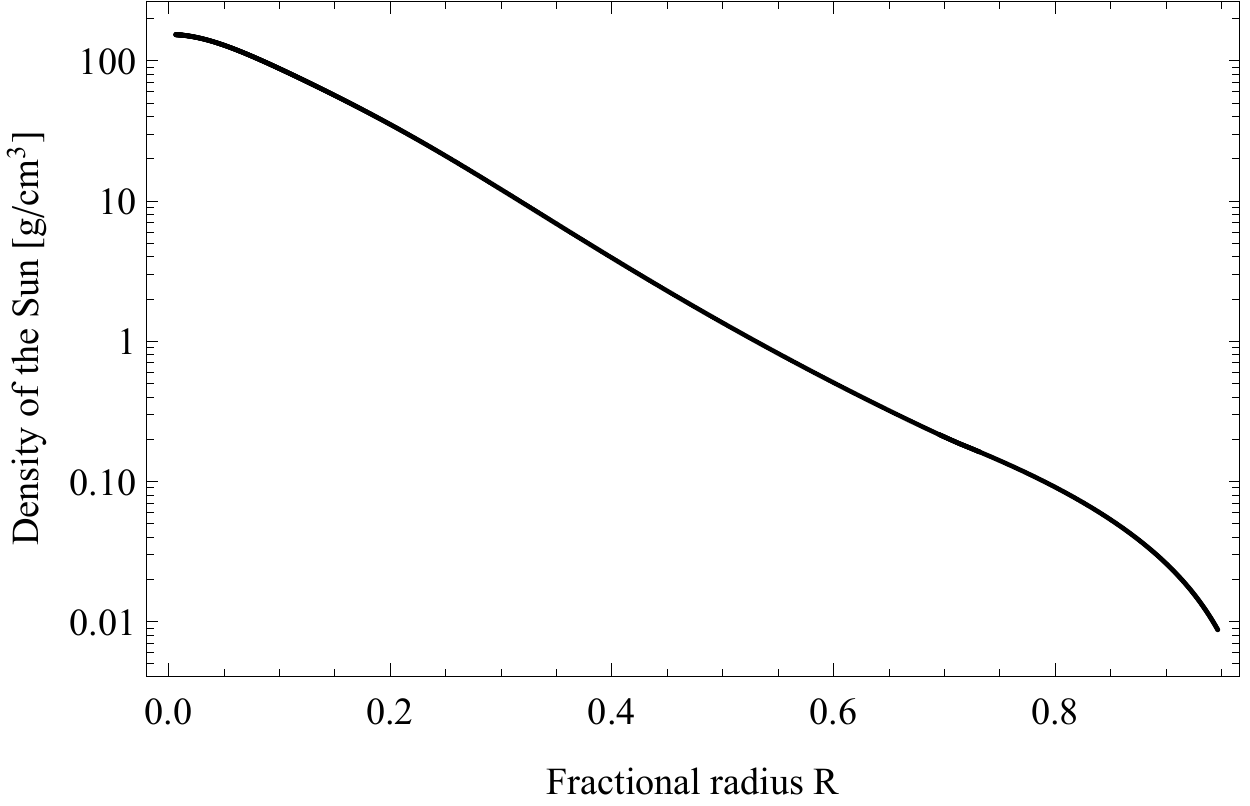}
	\caption{The solar density distribution function in terms of the fractional radius $R \equiv \frac{r}{R_{\odot}}$. This figure has been drawn by BP2004 data \cite{Bahcall:2004fg}.}
	\label{fig8}
\end{figure}

This model is specified by a runaway power-law (continuously decreasing) potential of the form
\begin{eqnarray}\label{eqnA1}
V(\phi)= M^{4+n}\phi^{-n},
\end{eqnarray}
where $n$ is a positive number, and $M$ is a parameter of mass scale. The chameleon scalar field's main feature is that its effective potential depends explicitly on the matter density.
\\By varying the action (\ref{eqn1}) with respect to the field, we obtain the following equation of motion
\begin{eqnarray}\label{eqnA2}
\square \phi = V_{,\phi} -  A^{3}(\phi) A_{,\phi}(\phi) \tilde{g}^{\mu\nu} \tilde{T}_{\mu\nu},
\end{eqnarray}
where $\tilde{T}_{\mu\nu} \equiv (-2/\sqrt{\tilde{g}}) \delta \mathcal{L}_m/\delta \tilde{g}_{\mu\nu}$ is the energy-momentum tensor, which is conserved in the Jordan frame
\begin{eqnarray}\label{eqnA3}
\tilde{\nabla}_\mu \tilde{T}^{\mu\nu} = 0.
\end{eqnarray}
From equation of state $\tilde{p} = \omega \tilde{\rho}$ we know the relationship between matter density in Einstein and Jordan frames \cite{Brax:2010kv}
\begin{eqnarray}\label{eqnA4}
\rho = A^{3(1 + \omega)}(\phi) \tilde{\rho}.
\end{eqnarray}
For non-relativistic matter, i.e. $\omega \approx 0$, we have $\tilde{T} = - \tilde{\rho} = - A^{-3}(\phi) \rho $. So the Eq.(\ref{eqnA2}) yields
\begin{eqnarray}\label{eqnA5}
\square \phi = V_{,\phi} + A_{,\phi} \rho,
\end{eqnarray}
where the right-hand-side of this equation can be written as the derivative of the effective potential $V_{\textsf{eff.}}(\phi) = V(\phi) + A(\phi) \rho$ with respect to the field $\phi$ where $\rho$ is the matter density.
The minimum of the potential is determined by using the equation $V_{\textsf{eff.},\phi}(\phi_{\textsf{min}}) = 0$, which leads to
\begin{eqnarray}\label{eqnA6}
\phi_{\textsf{min}} = \bigg[\frac{n M^{4+n} M_p}{\beta \rho}\bigg]^{\frac{1}{n+1}},
\end{eqnarray}
 Because the cosmological and local gravity experiments impose the condition $\frac{\beta \phi}{M_p} \ll 1$,  we assume that $e^{\frac{\beta \phi}{M_p}} \approx 1$. The effective mass of the field is given by
\begin{eqnarray}\label{eqnA7}
\begin{split}
&m^2_{\textsf{min}} \equiv V_{\textsf{eff.},\phi\phi} (\phi_{\textsf{min}})  = V_{,\phi\phi}(\phi_{\textsf{min}}) + \frac{\beta^2 \rho}{M^2_p} e^{\frac{\beta \phi_{\textsf{min}}}{M_p}} \\&~~~~~~~ = \frac{n(n+1) M^{4+n}}{\phi^{n+2}_{\textsf{min}}} + \frac{\rho \beta^2}{M_p^2},
\end{split}
\end{eqnarray}
where we have again used the assumption $e^{\frac{\beta \phi_{\textsf{min}}}{M_p}} \approx 1$.
\\To solve equation (\ref{eqnA5}) in a static spherically symmetric background we assume
\begin{eqnarray}\label{eqnA8}
\begin{split}
& \frac{d\phi}{dr} = 0  ~~~~~~~~~~~\textsf{at}~~~ r\longrightarrow 0,
\\& \phi \longrightarrow \phi_0 ~~~~~~~~~\textsf{at}~~~ r\longrightarrow \infty,
\end{split}
\end{eqnarray}
where the first condition is for non-singularity of the scalar field at the center of spherically symmetric body, while the second implies that the field
converges to a constant at infinity.
\\The solution to the Eq.(\ref{eqnA5}) can be obtained by expanding the field about its background as $\phi(r) = \phi_0 + \delta \phi$ up to linear order, where $\phi_0$ is the uniform background value and $\delta \phi$ is the perturbation induced by a spherically symmetric body like the Sun whose radius is
$R_{\odot}$ . Therefore the field equation turns into
\begin{eqnarray}\label{eqnA9}
\frac{d^2 \delta\phi}{dr^2} + \frac{2}{r} \frac{d\delta\phi}{dr}  =  m_{\textsf{min}}^2 (\phi_0) \delta\phi + \frac{\beta(\phi_0)}{M_p} \rho(r),
\end{eqnarray}
where the solution to this equation in dilute regions outside the body ($R:=\frac{r}{R_{\odot}}>1$) with constant density $\rho_0 \ll \bar{\rho}_{\odot}$ is given by
\begin{eqnarray}\label{eqnA10}
\delta \phi_{\textsf{out}}(R) = \delta\phi_{\textsf{in}}(1) \frac{1}{R} e^{-m_{\textsf{min}} R_{\odot} (R - 1)}, ~~~~~~~~~~~~~~~~~~(R>1)
\end{eqnarray}
where $\delta \phi_{\textsf{in}}(1)$ is the field value at the surface, i.e. $R=1$, coming from the continuity condition and $\bar{\rho}_{\odot}$ is the average density of the body. Adding this solution by the background value, we obtain the following solution for outside the body
\begin{eqnarray}\label{eqnA11}
\phi_{\textsf{out}}(R) = \phi_0 + \delta\phi_{\textsf{in}}(1) \frac{1}{R} e^{-m_{\textsf{min}} R_{\odot} (R - 1)}. ~~~~~~~~~~~~~~(R>1)
\end{eqnarray}
As can be seen from this, the scalar field induced by a celestial object, e.g. the Sun, is too small in large distances such that we can ignore its effects on another object in the solar system scale.
\\For inside the object, however, the density distribution of a spherical body like the Sun is a function of its fractional radius $R \equiv r / R_{\odot}$, as depicted in Fig.\ref{fig8}.

\begin{figure}[H]
	\begin{subfigure}{.5\textwidth}
		\centering
		\includegraphics[scale=0.59]{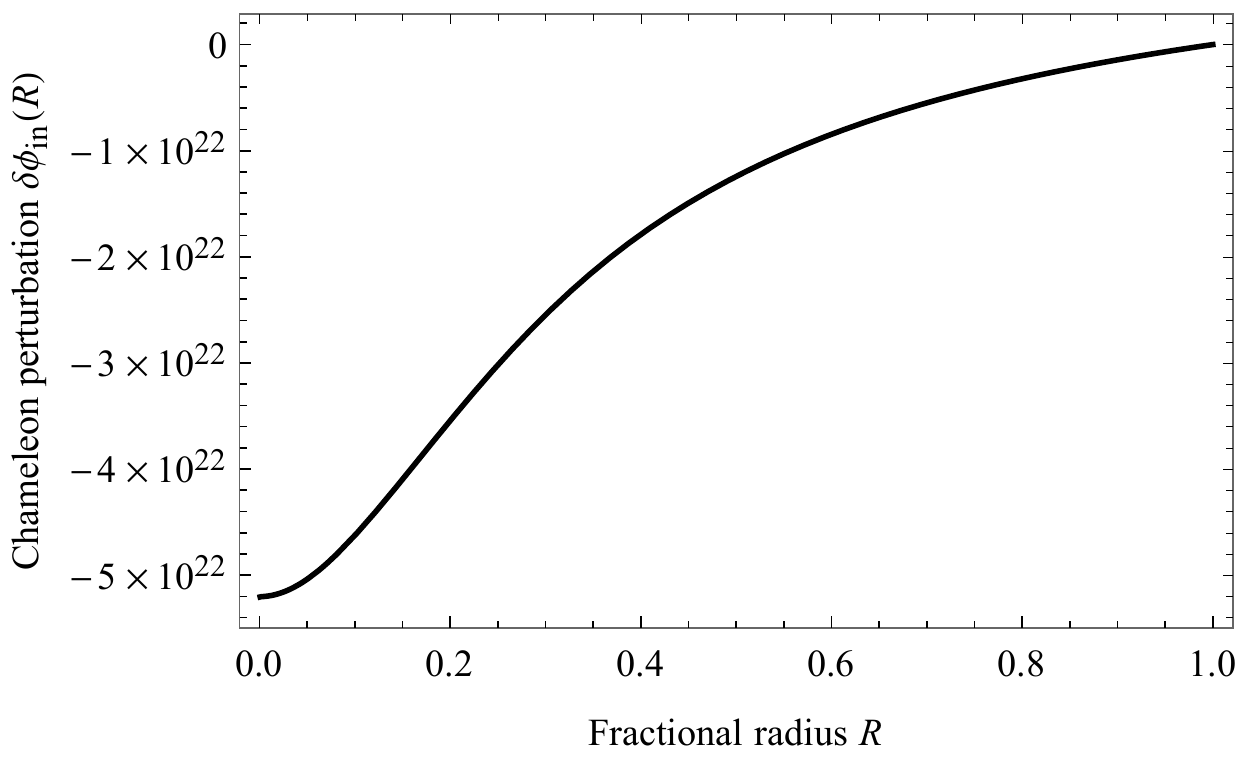}
		\caption{The perturbation $\delta \phi_{\textsf{in}}(R)$ in eV.}
		\label{fig9Left}
	\end{subfigure}
	\begin{subfigure}{.5\textwidth}
		\centering
		\includegraphics[scale=0.64]{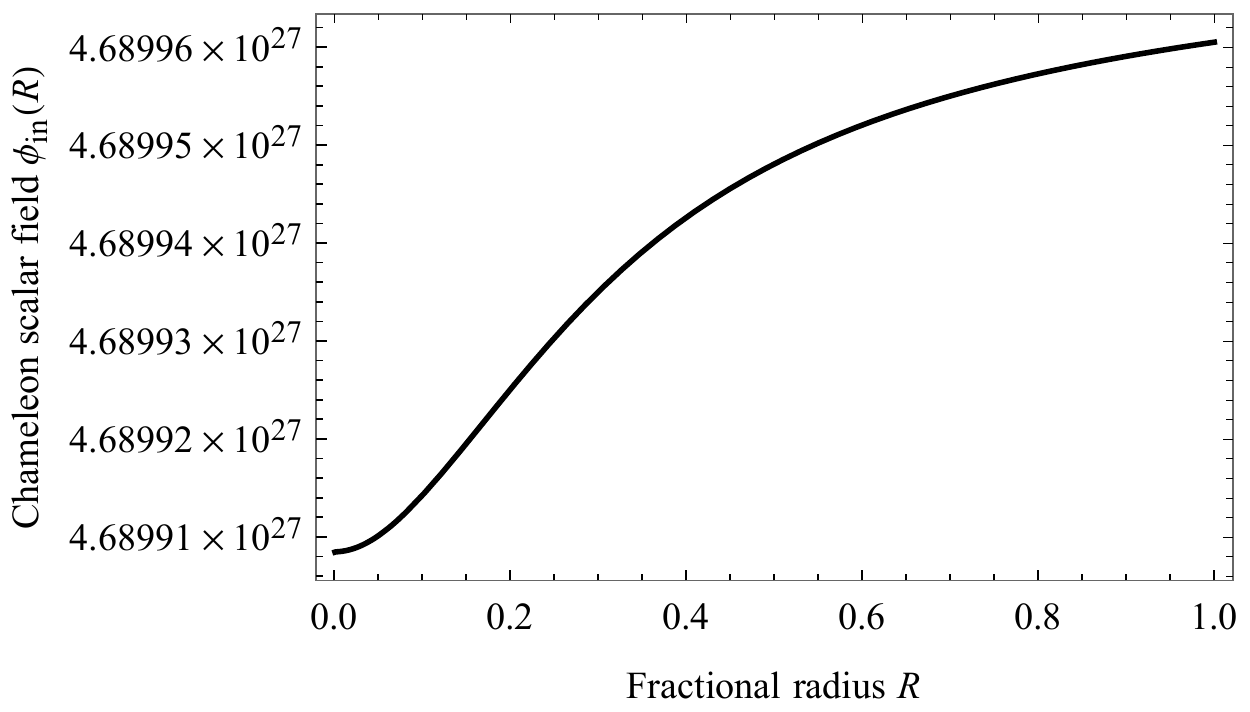}
		\caption{The chameleon inside the Sun in eV, $\phi_{\textsf{in}}(R) = \phi_0 + \delta \phi_{\textsf{in}}(R)$.}
		\label{fig9Right}
	\end{subfigure}
	\caption{The chameleon field profile for inside the Sun ($R\leq1$) with a $R$-dependent mass density distribution function. Note that we have assumed that $\beta \sim 1$, $n=1$, $\rho_0 \simeq 10^{-11}$eV$^{4}$ and $M \simeq 2$keV \cite{Waterhouse:2006wv}.}
	\label{fig9}
\end{figure}

\begin{figure}[H]
	\centering
	\includegraphics[scale=0.5]{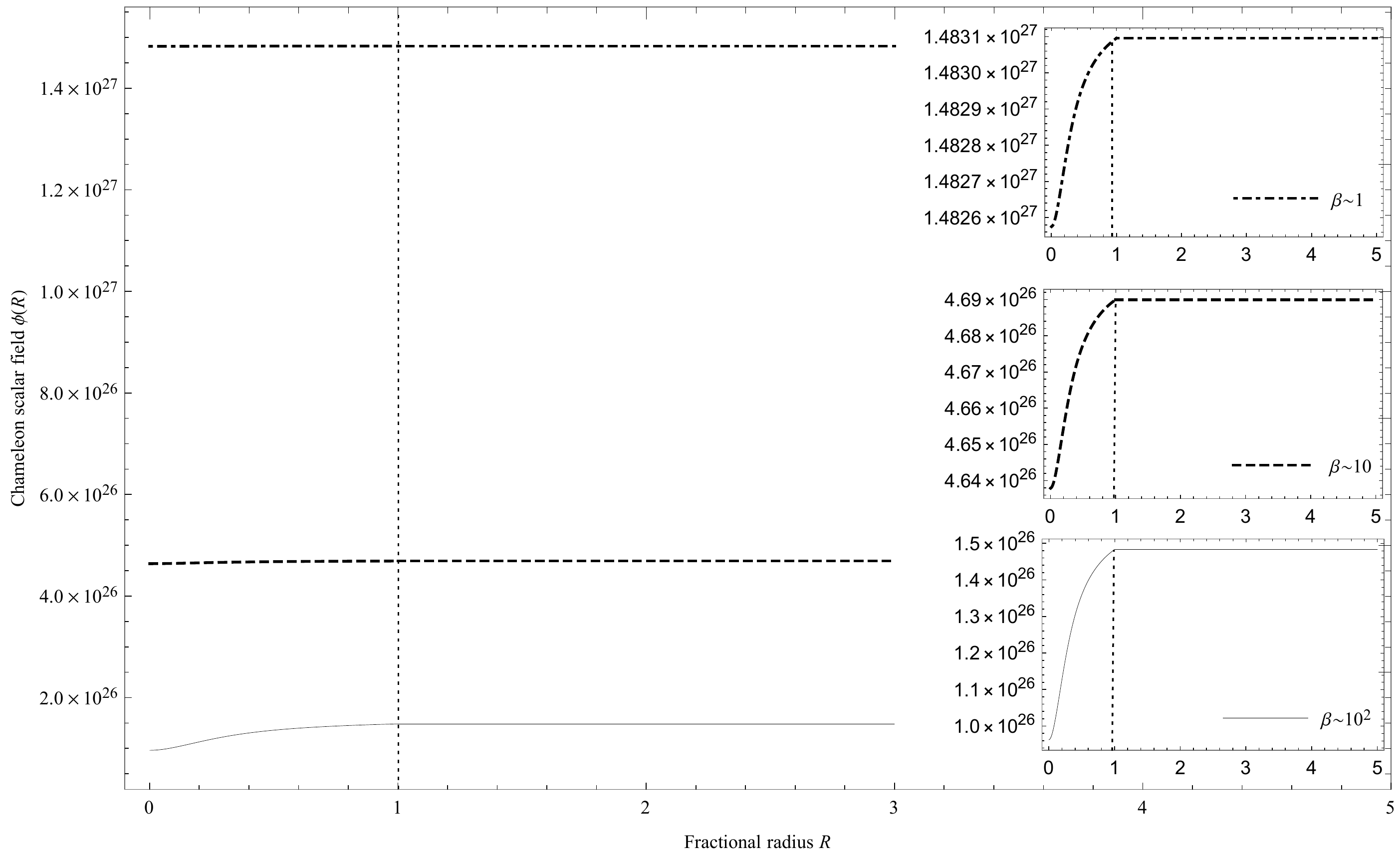}
	\caption{This figure shows how the coupling parameter might affect on the chameleon inside and outside the Sun. The chameleon in each case approaches to an asymptotic value outside the body, which increases with decreasing $\beta$. The field values are all in eV.}
	\label{fig10}
\end{figure}

With this density function, by solving numerically the equation (\ref{eqnA9}) we obtain an interpolating-function with a gravitation-strength coupling $\beta \sim 1$, as shown in Fig.\ref{fig9}. In this figure, both the resulting perturbation and the whole field inside the Sun are depicted.
We note that the resulting field profile can be sensitive to the change of the coupling $\beta$, see Fig.\ref{fig10}. This figure shows the behavior of the chameleon scalar field. We have used the value of $\delta\phi_{\textsf{in}}(R)$ at the surface of the Sun. The effects of the coupling parameter $\beta$ on the chameleon field for three different values of $\beta \in \{1,10,100\}$ are shown, implying that the chameleon tends to smaller asymptotic values when $\beta$ grows. We also note that the field's allowed range becomes smaller when $\beta$ reduces; hence, the field is being constant for $\beta \ll 1$.

\subsection{Symmetron mechanism}\label{subapp2}
In the symmetron screening mechanism, the screening is realized by symmetry restoration in sufficiently high-density regions in which $\phi = 0$ where the symmetron-matter coupling tends to zero. In low-density regions, the $\mathbb{Z}_2$-symmetry is spontaneously broken and $<\phi>\neq 0$.
An example of a $\mathbb{Z}_2$ symmetric coupling function and potential is:
\begin{eqnarray}\label{eqnA12}
\begin{split}
&A(\phi) = 1 + \frac{1}{2M^2} \phi^2+ \mathcal{O}\bigg(\frac{\phi^4}{M^4}\bigg),
\\&V(\phi) = V_0 - \frac{1}{2} \mu^2 \phi^2 + \frac{1}{4} \lambda \phi^4,
\end{split}
\end{eqnarray}
where $M$ and $\mu$ are two parameters of mass scale and $\lambda$ is a dimensionless  parameter. The equation of motion of the scalar field in a static spherically symmetric background is given by
\begin{eqnarray}\label{eqnA13}
\frac{d^2\phi}{dr^2} + \frac{2}{r} \frac{d\phi}{dr} = V_{\textsf{eff.},\phi} (\phi),
\end{eqnarray}
where the effective potential up to a constant is as follows
\begin{eqnarray}\label{eqnA14}
V_{\textsf{eff.}}(\phi) = \frac{1}{2M^2} (\rho - \rho_c) \phi^2 + \frac{1}{4} \lambda \phi^4,
\end{eqnarray}
where $\rho_c \equiv \mu^2 M^2$ is the critical density. The breaking or restoration of $\mathbb{Z}_2$-symmetry depends on whether the matter density is smaller or larger than the critical density. In the dilute regions the symmetry is spontaneously broken and the scalar field acquires a VEV (vacuum expectation value)
\begin{eqnarray}\label{eqnA15}
\phi_{\textsf{min,out}} = \pm \frac{\mu}{\sqrt{\lambda}} \sqrt{1-\frac{\rho}{\rho_c}} \approx \pm \frac{\mu}{\sqrt{\lambda}}. ~~~~~~~~~~~~~~~~~~~~ (\rho \ll \rho_c)
\end{eqnarray}
For the Sun, the scalar field effective mass $ m_{\textsf{min}}^2 \equiv \frac{d^2 V_{\textsf{eff.}}}{d\phi^2}|_{\phi_{\textsf{min}}}$ is then
\begin{eqnarray}\label{eqnA17}
\begin{split}
& m_{\textsf{out}} = \sqrt{2}\mu \sqrt{1 - \frac{\rho_{0}}{\rho_c}} \approx \sqrt{2} \mu ~~~~~~~~~~~~~~~~~~~~~(R>1)
\\& m_{\textsf{in}} = \mu \sqrt{\frac{\bar{\rho}_{\odot}}{\rho_c} - 1} \approx \mu \sqrt{\frac{\bar{\rho}_{\odot}}{\rho_c}}, ~~~~~~~~~~~~~~~~~~~~(R<1)
\end{split}
\end{eqnarray}
where as before $\rho_0$ is the background matter density and $\bar{\rho}_{\odot}$ is the solar average density and $R \equiv \frac{r}{R_{\odot}}$ .

\begin{figure}[H]
	\centering
	\includegraphics[scale=0.7]{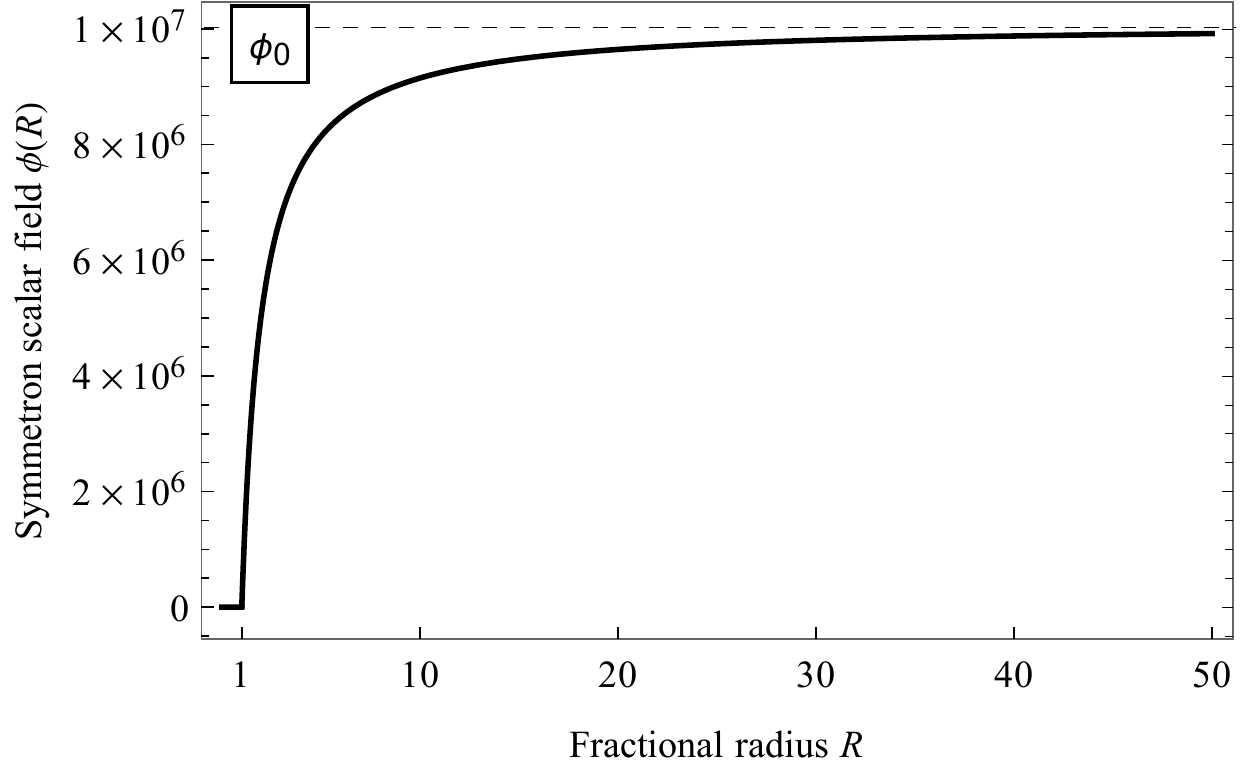}
	\caption{The symmetron field (in eV) of Eq.(\ref{eqnA26}) versus the fractional radius of the Sun.}
	\label{fig11}
\end{figure}

By expanding the scalar field around $\phi_{\textsf{min}}$ and keeping only the leading term we obtain the equation of motion
inside and outside a body like the Sun as follows
\begin{eqnarray}\label{eqnA21}
\frac{d^2 \delta\phi}{dR^2} + \frac{2}{R} \frac{d\delta\phi}{dR} =  R_{\odot}^2 m_{\textsf{in}}^2 (\delta\phi + \phi_0),~~~~~~~~~~~~~~~~~(R<1)
\end{eqnarray}
where we have used the approximation $V_{\textsf{eff.}}(\phi) \simeq \frac{\bar{\rho}_{\odot}}{M^2} \phi$ and $\phi_0 = \phi_{\textsf{min,out}}$ is the field value at the edge of the Sun \cite{Barreira:2015qxv}. The solution to the equation is
\begin{eqnarray}\label{eqnA22}
\delta\phi_{\textsf{in}}(R) = -\phi_0 + \frac{A}{R} \sinh\big((m_{\textsf{in}} R_{\odot}) R\big), ~~~~~~~~~~~~~(R<1)
\end{eqnarray}
$A$ is a constant to be determined by continuity conditions at the boundary.
For the outside, the equation of motion is given by
\begin{eqnarray}\label{eqnA23}
\frac{d^2 \delta\phi}{dR^2} + \frac{2}{R} \frac{d \delta\phi}{dR} = R_{\odot}^2 m_{\textsf{out}}^2 \delta\phi .~~~~~~~~~~~~~~~~~(R>1)
\end{eqnarray}
This equation has an analytical solution, which can be written as follows
\begin{eqnarray}\label{eqnA24}
\delta\phi_{\textsf{out}}(R) =  B \frac{e^{-( m_{\textsf{out}} R_{\odot}) R}}{R}, ~~~~~~~~~~~~~~~~~~~(R>1)
\end{eqnarray}
where $B$ is a constant. To specify the constants $A$ and $B$, we should use the continuity of $\delta\phi(R)$ and its first derivative at the $R=1$, which leads us to
\begin{eqnarray}\label{eqnA25}
\begin{split}
& A \longrightarrow  \phi_{0}~\frac{1 + m_{\textsf{out}} R_{\odot}}{m_{\textsf{in}} R_{\odot} \cosh (m_{\textsf{in}}R_{\odot}) + m_{\textsf{out}}R_{\odot} \sinh (m_{\textsf{in}} R_{\odot})} \\&
B \longrightarrow  - \phi_{0}~ e^{m_{\textsf{out}}R_{\odot}} \frac{m_{\textsf{in}} R_{\odot} \cosh (m_{\textsf{in}}R_{\odot}) - \sinh (m_{\textsf{in}} R_{\odot})}{m_{\textsf{in}} R_{\odot} \cosh (m_{\textsf{in}}R_{\odot}) + m_{\textsf{out}}R_{\odot} \sinh (m_{\textsf{in}} R_{\odot})}.
\end{split}
\end{eqnarray}
By assuming $m_{\textsf{in}} R_{\odot} \gg 1$ and $m_{\textsf{out}} R_{\odot} \ll 1$ and after some manipulation , the solution is derived as
\begin{eqnarray}\label{eqnA26}
\begin{split}
&\phi_{\textsf{in}}(R) = \frac{\phi_{0}}{m_{\textsf{in}} R_{\odot} \cosh (m_{\textsf{in}}R_{\odot})} \frac{\sinh\big((m_{\textsf{in}} R_{\odot}) R\big)}{R}  ~~~~~~~~~~~~~~~~~~~~~~~~~~~~(R<1) \\&
\phi_{\textsf{out}}(R) = \phi_{0} -  \phi_{0} \frac{m_{\textsf{in}} R_{\odot} - \tanh (m_{\textsf{in}} R_{\odot})}{ m_{\textsf{in}}R_{\odot}} \frac{e^{- m_{\textsf{out}} R_{\odot} (R-1)}}{R}. ~~~~~~~~~~~~~~~~(R>1)
\end{split}
\end{eqnarray}
In Fig. \ref{fig11}, we have plotted the symmetron scalar field as a function of the fractional radius $R$. Note that we have picked the numerical values $ M \simeq 10^{-4} M_{p}$, $\mu = 10^{-18}$eV, $\lambda$ $\sim 10^{-50}$ for model parameters. As the chameleon case, the symmetron tends to an asymptotic value $\phi_{0}$ at large distances from the Sun.


\begin{thebibliography}{99}

\bibitem{BP1}B. Pontecorvo, Sov. Phys. JETP {\bf 6}, 429 (1958) [Zh. Eksp. Teor. Fiz. {\bf 33},
549 (1958).
\bibitem{BP2} Z. Maki, M. Nakagawa and S. Sakata, Prog. Theor. Phys. {\bf 28}, 870 (1962).
\bibitem{Sun1}A. B. McDonald, “Nobel Lecture: The Sudbury Neutrino Observatory: Observation of flavor
change for solar neutrinos,” Rev. Mod. Phys. {\bf 88}, 030502 (2016).
\bibitem{Sun2}A. Bellerive et al. [SNO Collaboration], “The Sudbury Neutrino Observatory,” Nucl. Phys. B
{\bf 908}, 30 (2016), [arXiv:1602.02469 [nucl-ex]].
\bibitem{mu1}T. Kajita, “Nobel Lecture: Discovery of atmospheric neutrino oscillations,” Rev. Mod. Phys. {\bf 88}, 030501
(2016).
\bibitem{mu2} T. Kajita et al. [Super-Kamiokande Collaboration], Nucl. Phys. B {\bf 908}, 14 (2016).
\bibitem{MSWWolf}L. Wolfenstein, Phys. Rev. D {\bf 20}, 2634 (1979).
\bibitem{MSWMS}S. P. Mikheyev, A. Yu. Smirnov, Sov. J. Nucl. Phys. {\bf 42}, 913 (1985).
\bibitem{NSI}O. G. Miranda and H. Nunokawa, New J. Phys. {\bf 17}, 095002 (2015) [arXiv:1505.06254[hep-ph]].
\bibitem{NSI1}Shao-Feng Ge, S. J. Parke, Phys. Rev. Lett. 122, 211801 (2019) [arXiv:181208376[hep-ph]].
\bibitem{smir} S. F. Ge and A. Y. Smirnov, JHEP {\bf 1610}, 138 (2016), [arXiv:1607.08513 [hep-ph]].
\bibitem{NSIC1}L. Amendola, M. Baldi, and Ch. Wetterich, Phys. Rev. D {\bf 78}, 023015 (2008).
\bibitem{NSIC2} A.W. Brookfield, C. van de Bruck, D.F. Mota, D. Tocchini-Valentini, Phys. Rev. Lett. {\bf 96}, 061301 (2006), [arXiv:0503349 [astro-ph]].
\bibitem{NSIC3} M. C. González, Q. Liang, J. Sakstein, M. Trodden, arXiv:2011.09895 [astro-ph.CO].
\bibitem{NSIC4} S. Mandal, G. Y. Chitov, O. Avsajanishvili, B. Singha, T. Kahniashvili, arXiv:1911.06099 [hep-ph].
\bibitem{NSIC5} J.G. Salazar-Arias, A. Pérez-Lorenzana, Phys. Rev. D {\bf 101}, 083526 (2020), arXiv:1907.00131 [hep-ph].
\bibitem{NSIC6}E. A. Novikov, Amer.Res. J. Physics, {\bf 4}, 1 (2018).
\bibitem{MV11}R. Fardon, A. E. Nelson and N. Weiner, JCAP 0410, 005 (2004) [astro-ph/0309800].
\bibitem{MV12} P. Gu, X. Wang and X. Zhang, Phys. Rev. D {\bf 68}, 087301 (2003) [hep-ph/0307148].
\bibitem{MV13} C. Wetterich, Phys. Lett. B {\bf 655}, 201 (2007), arXiv:0706.4427 [hep-ph].
\bibitem{MV1}V. Barger, Patrick Huber, and D. Marfatia, Phys. Rev. Lett. {\bf 95}, 211802 (2005),[arXiv:hep-ph/0502196].
\bibitem{MV2}M.Cirelli, M.C. Gonzalez-Garcia, C. Pena-Garay, Nucl.Phys. B {\bf 719}, 219 (2005)
\bibitem{Nel}D. B. Kaplan, A. E. Nelson and N. Weiner, Phys. Rev. Lett. {\bf 93}, 091801 (2004) [hep-ph/0401099].
\bibitem{sad1}H. Mohseni Sadjadi, V. Anari, Phys. Rev. D {\bf 95}, 123521 (2017), [arXiv:1702.04244 [gr-qc]]
\bibitem{sad2}H. Mohseni Sadjadi, V. Anari, JCAP 10, 036 (2018),arXiv:1808.01903 [gr-qc]
\bibitem{sam}M. Sami, Sh. Myrzakul, M. Al Ajmi,  Phys.Dark Univ. {\bf 30}, 100675 (2020) 100675, [arXiv:1912.12026 [gr-qc]]
\bibitem{amen}L. Amendola, Sh. Tsujikawa, ``Dark Energy Theory and Observations'' (Cambridge University Press, Cambridge, UK, 2010).
\bibitem{mat1} R. Bean, E. E. Flanagan, and M. Trodden, Phys. Rev. D {\bf 78}, 023009 (2008).
\bibitem{mat2} H. Mohseni Sadjadi, M. Honardoost, H. R. Sepangi, Phys. Dark Univ. {\bf 14}, 40 (2016), [arXiv:1504.05678 [gr-qc]].
 \bibitem{mat3}M. Honardoost, H. Mohseni Sadjadi, H. R. Sepangi, Gen Relativ Gravit {\bf 48}, 125 (2016)[arXiv:1508.06022 [gr-qc]]
\bibitem{Khoury:2003rn}
J.~Khoury and A.~Weltman,
Phys.\ Rev.\ D {\bf 69}, 044026 (2004), [arXiv:[astro-ph/0309411]].
\bibitem{Waterhouse:2006wv}
T.~P.~Waterhouse, [arXiv:[astro-ph/0611816]].
\bibitem{Burrage:2017qrf}
C.~Burrage and J.~Sakstein, Living Rev.\ Rel.\  {\bf 21}, no. 1, 1 (2018).
[arXiv:1709.09071 [astro-ph.CO]].
\bibitem{Tsujikawa:2009yf}
S.~Tsujikawa, T.~Tamaki and R.~Tavakol,
JCAP {\bf 0905}, 020 (2009), [arXiv:0901.3226 [gr-qc]].
\bibitem{Hinterbichler:2011ca}
K.~Hinterbichler, J.~Khoury, A.~Levy and A.~Matas,
Phys. Rev. D \textbf{84}, 103521 (2011)
[arXiv:1107.2112 [astro-ph.CO]]
\bibitem{sad3} H. Mohseni Sadjadi, JCAP 01,031(2017), [arXiv:1609.04292 [gr-qc]]
\bibitem{Mix}M. Lindner, T. Ohlsson, and W. Winter, Nucl. Phys. B 607 (2001), 326, [hep-ph/0103170].
\bibitem{Aharmim:2018fme}
B.~Aharmim \textit{et al.} [SNO],Phys. Rev. D \textbf{99}, no.3, 032013 (2019)[arXiv:1812.01088 [hep-ex]].
\bibitem{NeutLifTime}
S.~Choubey, D.~Dutta and D.~Pramanik, JHEP \textbf{08}, 141 (2018) [arXiv:1805.01848 [hep-ph]];
M.~Escudero and M.~Fairbairn, Phys. Rev. D \textbf{100}, no.10, 103531 (2019)[arXiv:1907.05425 [hep-ph]].
\bibitem{Zyla:2020zbs}
P.~A.~Zyla \textit{et al.} [Particle Data Group], PTEP \textbf{2020}, no.8, 083C01 (2020).
\bibitem{Agostini:2018uly}
M.~Agostini \textit{et al.} [BOREXINO],
Nature \textbf{562}, no.7728, 505-510 (2018).
\bibitem{Faraoni:1998qx}
V.~Faraoni, E.~Gunzig and P.~Nardone,
Fund.\ Cosmic Phys.\  {\bf 20}, 121 (1999).
[gr-qc/9811047].
\bibitem{Carneiro:2004rt}
D.~F.~Carneiro, E.~A.~Freiras, B.~Goncalves, A.~G.~de Lima and I.~L.~Shapiro,
Grav.\ Cosmol.\  {\bf 10}, 305 (2004).
[gr-qc/0412113].
\bibitem{Bean} R. Bean, E. E. Flanagan, and M. Trodden, Phys. Rev. D 78, 023009
(2008)
\bibitem{Cottingham:2007zz}
W.~N.~Cottingham and D.~A.~Greenwood,
``An introduction to the standard model of particle physics,''(Cambridge Univ. Press, 1998).
\bibitem{MohseniSadjadi:2017jne}
H.~Mohseni Sadjadi and A.~P.~Khosravi,
JCAP {\bf 1804}, 008 (2018), [arXiv:1711.06607 [hep-ph]].
\bibitem{Cardall:1996cd}
C.~Y.~Cardall and G.~M.~Fuller,
Phys.\ Rev.\ D {\bf 55}, 7960 (1997), [hep-ph/9610494].
\bibitem{Visinelli:2015uva}
L.~Visinelli, Gen.\ Rel.\ Grav.\  {\bf 47}, no. 5, 62 (2015),
[arXiv:1410.1523 [gr-qc]].
\bibitem{Buoninfante:2019der}
L.~Buoninfante, G.~G.~Luciano, L.~Petruzziello and L.~Smaldone,Phys. Rev. D \textbf{101}, no.2, 024016 (2020) [arXiv:1906.03131 [gr-qc]].
\bibitem{Chakraborty:2015vla}
S.~Chakraborty, JCAP \textbf{10}, 019 (2015) [arXiv:1506.02647 [gr-qc]].
\bibitem{Barger:1998xk}
V.~D.~Barger, J.~G.~Learned, S.~Pakvasa and T.~J.~Weiler,
Phys.\ Rev.\ Lett.\  {\bf 82}, 2640 (1999), [astro-ph/9810121].
\bibitem{Blennow:2005yk}
M.~Blennow, T.~Ohlsson and W.~Winter, JHEP {\bf 0506}, 049 (2005),[hep-ph/0502147].
\bibitem{GonzalezGarcia:2007ib}
M.~C.~Gonzalez-Garcia and M.~Maltoni,
Phys.\ Rept.\  {\bf 460}, 1 (2008),[arXiv:0704.1800 [hep-ph]].
\bibitem{Joshipura:2002fb}
A.~S.~Joshipura, E.~Masso and S.~Mohanty,
Phys.\ Rev.\ D {\bf 66}, 113008 (2002),[hep-ph/0203181].
\bibitem{WEPtest}
J.~O.~Dickey, P.~L.~Bender, J.~E.~Faller, X.~X.~Newhall, R.~L.~Ricklefs, J.~G.~Ries, P.~J.~Shelus, C.~Veillet, A.~L.~Whipple and J.~R.~Wiant, \textit{et al.} Science \textbf{265}, 482-490 (1994); J.~G.~Williams, S.~G.~Turyshev and D.~H.~Boggs, Int. J. Mod. Phys. D \textbf{18}, 1129-1175 (2009); J.~G.~Williams, S.~G.~Turyshev and D.~Boggs, Class. Quant. Grav. \textbf{29}, 184004 (2012); T.~A.~Wagner, S.~Schlamminger, J.~H.~Gundlach and E.~G.~Adelberger, Class. Quant. Grav. \textbf{29}, 184002 (2012).
\bibitem{Adey:2018zwh}
D.~Adey \textit{et al.} [Daya Bay], Phys. Rev. Lett. \textbf{121}, no.24, 241805 (2018) [arXiv:1809.02261 [hep-ex]].
\bibitem{Bak:2018ydk}
G.~Bak \textit{et al.} [RENO], Phys. Rev. Lett. \textbf{121}, no.20, 201801 (2018) [arXiv:1806.00248 [hep-ex]].
\bibitem{Beacom:2002cb}
J.~F.~Beacom and N.~F.~Bell, Phys. Rev. D \textbf{65}, 113009 (2002) [arXiv:hep-ph/0204111 [hep-ph]].
\bibitem{Aharmim:2011vm}
B.~Aharmim \textit{et al.} [SNO], Phys. Rev. C \textbf{88}, 025501 (2013) [arXiv:1109.0763 [nucl-ex]].
\bibitem{Parke:1986jy}
S.~J.~Parke, Phys. Rev. Lett. \textbf{57}, 1275-1278 (1986).
\bibitem{Petcov:1987zj}
S.~T.~Petcov, Phys. Lett. B \textbf{200}, 373-379 (1988).
\bibitem{Bandyopadhyay:2001ct}
A.~Bandyopadhyay, S.~Choubey and S.~Goswami,
Phys.\ Rev.\ D {\bf 63}, 113019 (2001), [hep-ph/0101273].
\bibitem{Strumia:2006db}
A.~Strumia and F.~Vissani,``Neutrino masses and mixings and...,'' [arXiv:hep-ph/0606054 [hep-ph]].
\bibitem{Gando:2013nba}
A.~Gando \textit{et al.} [KamLAND], Phys. Rev. D \textbf{88}, no.3, 033001 (2013) [arXiv:1303.4667 [hep-ex]].
\bibitem{SK}
J.~Hosaka \textit{et al.} [Super-Kamiokande], Phys. Rev. D \textbf{73}, 112001 (2006) [arXiv:hep-ex/0508053 [hep-ex]];
J.~P.~Cravens \textit{et al.} [Super-Kamiokande], Phys. Rev. D \textbf{78}, 032002 (2008) [arXiv:0803.4312 [hep-ex]];
K.~Abe \textit{et al.} [Super-Kamiokande], Phys. Rev. D \textbf{83}, 052010 (2011) [arXiv:1010.0118 [hep-ex]];
K.~Abe \textit{et al.} [Super-Kamiokande], Phys. Rev. D \textbf{94}, no.5, 052010 (2016) [arXiv:1606.07538 [hep-ex]].
\bibitem{Smirnov:2016xzf}
A.~Y.~Smirnov,``Solar neutrinos: Oscillations or No-oscillations?,'' [arXiv:1609.02386 [hep-ph]].
\bibitem{Guidry:2018ocm}
M. Guidry and J. Billings, [arXiv:1812.00035 [astro-ph.SR]].
\bibitem{Esteban:2020cvm}
I.~Esteban, M.~C.~Gonzalez-Garcia, M.~Maltoni, T.~Schwetz and A.~Zhou, JHEP \textbf{09}, 178 (2020) [arXiv:2007.14792 [hep-ph]]; see also \href{http://www.nu-fit.org/}{www.nu-fit.org}.
\bibitem{Khoury:2003aq}
J.~Khoury and A.~Weltman,
Phys. Rev. Lett. \textbf{93}, 171104 (2004)
[arXiv:astro-ph/0309300 [astro-ph]].
\bibitem{Mota:2006fz}
D.~F.~Mota and D.~J.~Shaw, Phys. Rev. D \textbf{75}, 063501 (2007)
[arXiv:hep-ph/0608078 [hep-ph]].
\bibitem{Touboul:2008zz}
P.~Touboul, Space Sci. Rev. \textbf{148}, 455-474 (2008).
\bibitem{Mester:2001}
J.~Mester~et al., Classical Quantum Gravity~\textbf{18}, 2475 (2001).
\bibitem{Nobili:2000}
A.M.~Nobili~et al.,  Classical Quantum Gravity~\textbf{17}, 2347 (2000).
\bibitem{Hinterbichler:2010es}
K.~Hinterbichler and J.~Khoury, Phys. Rev. Lett. \textbf{104}, 231301 (2010), [arXiv:1001.4525 [hep-th]].
\bibitem{Baerwald:2012kc}
P.~Baerwald, M.~Bustamante and W.~Winter, JCAP {\bf 1210}, 020 (2012), [arXiv:1208.4600 [astro-ph.CO]].
\bibitem{Bahcall:2004pz}
J.~N.~Bahcall, A.~M.~Serenelli and S.~Basu, Astrophys. J. Lett. \textbf{621}, L85-L88 (2005),[arXiv:astro-ph/0412440 [astro-ph]].
\bibitem{Drouin:2011pna}
P.~L.~Drouin,
``Three-Phase Extraction of the Electron Neutrino Survival Probability at the Sudbury Neutrino Observatory'', Ph.D. Thesis, Carleton University (2011); For further
see \href{http://falcon.phy.queensu.ca/SNO/sno/publications.html}{http://falcon.phy.queensu.ca/SNO/sno/publications.html}..
\bibitem{Spergel:2006hy}
D.~N.~Spergel {\it et al.} Astrophys.\ J.\ Suppl.\  {\bf 170}, 377 (2007),[astro-ph/0603449].
\bibitem{Upadhye:2012qu}
A.~Upadhye, Phys.\ Rev.\ D {\bf 86}, 102003 (2012), [arXiv:1209.0211 [hep-ph]].
\bibitem{SKnonstand}
G.~Mitsuka \textit{et al.} [Super-Kamiokande], Phys. Rev. D \textbf{84}, 113008 (2011) [arXiv:1109.1889 [hep-ex]].
\bibitem{HK}
K.~Abe \textit{et al.} [Hyper-Kamiokande], [arXiv:1805.04163 [physics.ins-det]].
\bibitem{JUNO}
F.~An \textit{et al.} [JUNO], J. Phys. G \textbf{43}, no.3, 030401 (2016) [arXiv:1507.05613 [physics.ins-det]].
\bibitem{Liao:2017awz}
J.~Liao, D.~Marfatia and K.~Whisnant, Phys. Lett. B \textbf{771}, 247-253 (2017) [arXiv:1704.04711 [hep-ph]].
\bibitem{DUNE}
R.~Acciarri \textit{et al.} [DUNE], [arXiv:1512.06148 [physics.ins-det]].
\bibitem{Liao:2016orc}
J.~Liao, D.~Marfatia and K.~Whisnant, JHEP \textbf{01}, 071 (2017) [arXiv:1612.01443 [hep-ph]].
\bibitem{DUNEnonstand}
B.~Abi \textit{et al.} [DUNE],``Prospects for Beyond the Standard Model Physics Searches at the Deep Underground Neutrino Experiment,'' [arXiv:2008.12769 [hep-ex]].
\bibitem{ErthDen}
A.~M.~Dziewonski and D.~L.~Anderson, Phys. Earth Planet. Interiors \textbf{25}, 297-356 (1981);
K. Hoshina and H. Tanaka, EGU General Assembly Conference Abstracts \textbf{14}, 3246 (2012).
\bibitem{Choubey:2017dyu}
S.~Choubey, S.~Goswami and D.~Pramanik, JHEP \textbf{02}, 055 (2018) [arXiv:1705.05820 [hep-ph]].
\bibitem{Brax:2010kv}
P.~Brax, C.~van de Bruck, D.~F.~Mota, N.~J.~Nunes and H.~A.~Winther,
Phys. Rev. D \textbf{82}, 083503 (2010), [arXiv:1006.2796 [astro-ph.CO]].
\bibitem{Bahcall:2004fg}
J.~N.~Bahcall and M.~H.~Pinsonneault,
Phys.\ Rev.\ Lett.\  {\bf 92}, 121301 (2004), [astro-ph/0402114]
\bibitem{Barreira:2015qxv}
A.~M.~R.~Barreira,
``Structure formation in modified gravity cosmologies'', (Springer, New York, 2016), pp. 1-22.





\end{thebibliography}
\end{document}